\documentclass{sig-alternate}

\newdef{mydef}{Definition}
\newdef{myproblem}{Problem}

\usepackage{flushend}
\usepackage[caption=false]{subfig}
\usepackage{multirow}
\usepackage{paralist}
\usepackage{amsmath}
\usepackage{amssymb}

\begin{document}
%

\title{Microscopic Evolution of Social Networks \\by Triad Position Profile}

\numberofauthors{3} 
%
\author{
%
%
\alignauthor
Yang Yang\\
       \affaddr{Department of Computer Science and Engineering}\\
       \affaddr{University of Notre Dame}\\
       \email{yyang1@nd.edu}
\alignauthor
Yuxiao Dong\\
       \affaddr{Department of Computer Science and Engineering}\\
       \affaddr{University of Notre Dame}\\
       \email{ydong1@nd.edu}
\alignauthor
Nitesh V. Chawla\\
       \affaddr{Department of Computer Science and Engineering}\\
       \affaddr{University of Notre Dame}\\
       \email{nchawla@nd.edu}
}
\date{01 Oct 2013}

\maketitle

\begin{abstract}
Disentangling the mechanisms underlying the social network evolution is one of social science's unsolved puzzles. Preferential attachment is a powerful mechanism explaining social network dynamics, yet not able to explain all scaling-laws in social networks. Recent advances in understanding social network dynamics demonstrate that several scaling-laws in social networks follow as natural consequences of triadic closure. Macroscopic comparisons between them are discussed empirically in many works. However the network evolution drives not only the emergence of macroscopic scaling but also the microscopic behaviors. Here we exploit two fundamental aspects of the network microscopic evolution: the individual influence evolution and the process of link formation. First we develop a novel framework for the microscopic evolution, where the mechanisms of preferential attachment and triadic closure are well balanced. Then on four real-world datasets we apply our approach for two microscopic problems: node's prominence prediction and link prediction, where our method yields significant predictive improvement over baseline solutions. Finally to be rigorous and comprehensive, we further observe that our framework has a stronger generalization capacity across different kinds of social networks for two microscopic prediction problems. We unveil the significant factors with a greater degree of precision than has heretofore been possible, and shed new light on networks evolution.

\end{abstract}


\section{Introduction}
\label{sec_intro}
Disentangling the mechanisms underlying the social network evolution is one of social science's unsolved puzzles. Recent advances in research bring us a wide variety of principles and models for the growth of complex networks. In most works these principles/models are validated from perspectives of macroscopic scaling-laws, such as {\it power-law} degree distribution \cite{acm:preferential}, {\it attachment kernel} \cite{acm:attachmentkernel} and {\it clustering coefficient} as function of node degree \cite{acm:preferential}. Anecdotal evidence that preferential attachment is a powerful mechanism underlying the emergence of scale-free property in social networks, where new links are established preferentially to more popular nodes in a network, is ubiquitous. However it is also evident that the {\it preferential attachment} principle is not able to explain all scaling laws \cite{acm:newman2001} \cite{acm:li2010} \cite{acm:triad1}. Further study in \cite{acm:newman2001} and \cite{acm:li2010} shows that an individual's link formation significantly relies on its neighbors. The principle of {\it triadic closure} has been empirically demonstrated to be relevant for above three macroscopic scaling laws in the work of \cite{acm:jure2008} \cite{acm:validation1} \cite{acm:validation2} \cite{acm:validation3} \cite{acm:triad1}, expressly or implicitly. To summarize, both preferential attachment and triadic closure are strong force shaping the network dynamics. The questions is whether they can be balanced and unified in one framework.

Tremendous works have been proposed to compare principles in macroscopic level. However the network evolution drives not only the emergence of macroscopic scaling but also the microscopic behavior. Different from prior research we exploit distinctness of principles from the {\it microscopic} perspective. The evolution of social network affects individuals in two aspects: 1) nodal influence varies over time; 2) new links are attached to existing nodes. They are highly intertwined. Formation of new links will lead to enhancing a node's influence or prominence, and the increment of node's influence over time will attract more links ({\it{Preferential Attachment} \cite{acm:preferential}}). Consider Twitter as an example: as an individual rises in prominence, he/she generates more followers or links. Likewise a website grows in prominence on the basis of its connections ({\it PageRank} \cite{acm:pagerank}). {\it We posit a richer framework in the network evolution analysis and modeling should be capable of describing both of the influence evolution and the link formation mechanism. This is the central theme of our paper.}

Influence analysis and modeling is a subject focus in social networks. This includes influence maximization  
\cite{acm:influencemaximize0} \cite{acm:influencemaximize1}, influence selection and quantification \cite{acm:goyal}, and influence validation \cite{acm:feedback}. In addition, different influence models \cite{acm:socialaction1} \cite{acm:socialaction2} and centrality measures, such as Pagerank \cite{acm:pagerank}, Betweenness, \cite{acm:betweenness} Closeness \cite{acm:closeness}, and Clustering Coefficient \cite{acm:clusteringcoefficient} have been used for discovering influential nodes in a network. These methods are limited as they are not predictive about possible rise to prominence or influence of a node in future, and are also not consistent in their performances across different types of networks. Thus, a fundamental question that we consider in this paper:  {\it{is there a generic approach for the influence analysis and prediction of the prominence of a node?}} 

As mentioned, the process of link formation is also an integrated aspect of network evolution. In recent work \cite{acm:nature}, a measure of attractiveness that balanced popularity (i.e., {\it preferential attachment}) and similarity (i.e., {\it common neighbors}) was shown to have a better interpretation of the link formation mechanism. Additionally in the work of Liben-Nowell and Kleinberg \cite{acm:linkprediction1} and Lichtenwalter et al. \cite{acm:linkprediction2} it was observed that there is no single feature that is capable of outperforming uniformly in different networks, and \cite{acm:linkprediction2} also developed a supervised learning method that included the different features and outperformed the singular features. While there is a body of work in link prediction \cite{acm:linkprediction3} \cite{acm:linkprediction4} \cite{acm:linkprediction6} \cite{acm:dong}, there is a paucity of an understanding of the evolutionary processes that guide link formation. A fundamental question that we consider here is: {\it{how to develop a coherent model that captures influence evolution to inform link prediction?}} 

Modeling social networks serves to help us understand how social networks form and evolve. Besides providing approaches for concrete problems, we are more interested to know what is the fundamental principle in social network evolution. To further study, we ask the question, whether the models developed are transferable from one network to another. That is: is the model for prominence prediction and link prediction, generic enough to learn from one social network and make a prediction on another social network? With these rigorous analysis we unveil that {\it triadic closure} could be identified as one of the fundamental principles in the social network evolution. Our contributions are summarized as follows:
\begin{itemize}
\item In Section~\ref{sec_triad_measure} we discuss two popular principles (preferential attachment and triadic closure) and their consequences on the microscopic evolution of social networks. We develop a framework called {\bf triad position profile} where the trade-offs between two principles are optimized.
\item In Section~\ref{sec_infer_prominence} and ~\ref{sec_link_prediction} we apply our approach for node's prominence prediction and link prediction. We validate that our framework can interpret the individual influence evolution and the link formation mechanism better than has heretofore been possible.
\item In Section~\ref{sec_generalization} the validity of generality is tested on four real-world networks, which demonstrates that our methodology has a better interpretation of mechanisms underlying network evolution.
\end{itemize}

Overall our work provides microscopic insights about social network evolution with applications ranging from link prediction to inferring the future prominence of an individual node.

\section{Preliminaries}
\label{sec_preliminaries}
\subsection{Datasets}
\label{sec_dataset}
In this paper we examine our approaches and perform our analysis on four social networks. The {\bf Condmat} network \cite{acm:linkprediction2} is extracted from a stream of 19,464 multi-agent events representing condensed matter physics collaborations from 1995 to 2000. Based on the {\bf DBLP} dataset from \cite{acm:dblp} we attach timestamps for each collaboration and choose 3,215 authors who published at least 5 papers. {\bf Enron} dataset \cite{acm:enron} contains information of email communication among 16,922 employees in Enron Corporate from 2001.1.1 to 2002.3.31. The {\bf Facebook} dataset is used by Viswanath et al. \cite{acm:facebook}, which contains wall-to-wall post relationship among 11,470 users between 2004.10 and 2009.1.

\subsection{Problem Definition}
\label{sec_pareto}
Network evolution is usually reflected in changes of node's prominence and new links formed with the other nodes.  First, the social network evolution impacts the prominence (social status) of node; in addition, the node also affects its local neighbourhood and beyond (via link formation or link dissolution). In order to give insights into the network dynamics, we provide several definitions and formulate two concrete problems for the ease of evaluation and comparison.

\begin{figure}[t]
	\centering
	\subfloat[Degree Ranking]{\includegraphics[width=1.3in]{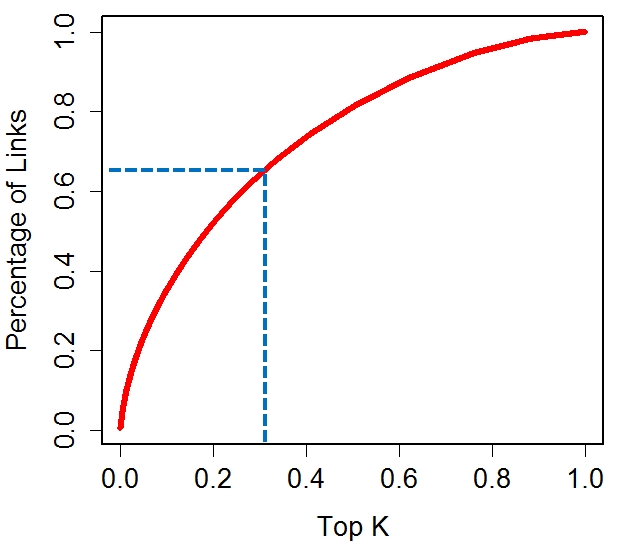}}\quad
	\subfloat[PageRank Ranking]{\includegraphics[width=1.3in]{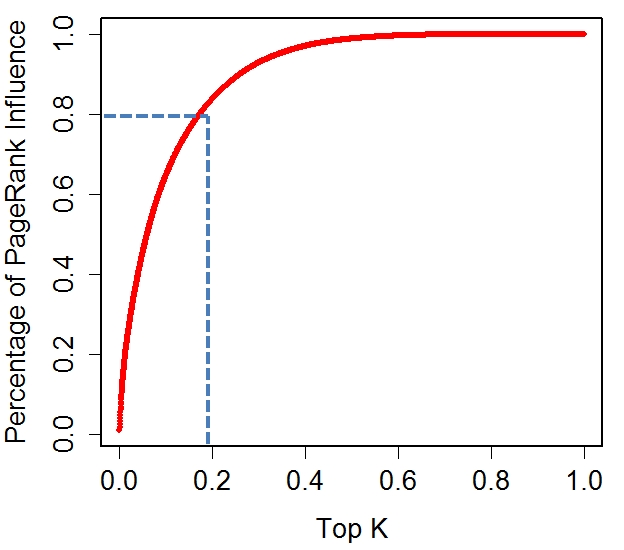}}\\	
	\caption{Pareto Principle.}
\label{fig_pareto_principle}
\end{figure}

On a global level, influential nodes or prominent nodes have intrinsically higher strength of influence than others due to the network topology. Through our study, we have found that a small number of nodes occupy large portion of network resources. For example, in Figure~\ref{fig_pareto_principle}(b) top 20\% (ranked by {PageRank}) nodes occupy about 80\% {\it PageRank} influence in DBLP network. This satisfies {\it Pareto Principle} (also known as 80-20 rule) \cite{acm:pareto}. To better understand and model the effects of network evolution on node's prominence, we partition nodes into two sets {\it prominent nodes} and {\it non-prominent nodes}.
Based on {\it Pareto Principle}, their definitions are given as follows:
\begin{mydef}{\bf Prominent Node}
In a network $G=(V, E)$ a node $v$ is a {\it prominent node} under influence measurement $\mathbb{M}$ if and only if $ \frac{| \{ u | \mathbb{M}(u) \leq \mathbb{M}(v) \}|}{|V|} \geq 0.8$.
\end{mydef}
\begin{mydef}{\bf Non-Prominent Node}
In a network $G=(V, E)$, a node $v$ is a {\it non-prominent node} under influence measurement $\mathbb{M}$ if and only if $ \frac{| \{ u | \mathbb{M}(u) > \mathbb{M}(v) \}|}{|V|} \geq 0.2$.
\end{mydef}
In following sections we denote the set of {\it prominent nodes} as ${\bf PN}$ and the set of {\it non-prominent nodes} as ${\bf NPN}$.

As we postulated, an arguably generic approach for the network evolution analysis should have the predictability of a node's prominence in the future. Therefore we formulate a concrete task, the prominence prediction problem, where we can directly evaluate different approaches and facilitate our findings of the underlying principles.

While the link formation is a closely intertwined process with the changes of node's prominence, which is a negligible part in our analysis. To validate the {\it link formation} mechanism, we employ the link prediction problem as our evaluation metric. The associated definitions are as follows:
\begin{mydef} {\bf Time-varying Network}
The time-varying network at time $t$ is denoted as $G_{t}=(V, E, T_{V}, T_{E})$, where $V$ is the set of nodes and $E$ is the set of links among nodes, $T_{V}$ is the set of arriving time of all nodes and $T_{E}$ is the time log of all links.
\end{mydef}
\begin{myproblem} {\bf Prominence Prediction}
In a time-varying network $G_{t}=(V, E, T_{V}, T_{E})$ the prominence prediction task is, for the set of nodes $ V_{t } = \{ v | v \in V, T_{V} (v) = t, v \notin {\bf PN}_{t} \}$, where ${\bf PN}_{t}$ is the set of {\it prominent nodes} measured in network $G_{t}$. How reliably can we infer whether a node $v$ ($v \in V_{t}$) will belong to the set of ${\bf PN}_{t+\Delta T}$?
\label{def_prominence_prediction}
\end{myproblem}
In order to demonstrate the discrimination of different principles, $\Delta T$ is selected large enough for the node influence evolution.
\begin{myproblem} {\bf Link Prediction}
In a time-varying network $G_{t}=(V, E, T_{V}, T_{E})$, the link prediction task in such network is to predict whether there will be a link between a pair of nodes $u$ and $v$ at time $t+\Delta T$, where $u, v \in V$ and $e(u, v) \notin E$.
\end{myproblem}

These concrete problems provide us quantitative and microscopic views of network evolution, which also make it convenient for principles comparison.

Besides verifying the generality of approaches across two intimately interacted processes of network evolution, we also study whether the learned predictors can generalize across different domains of social networks for both problems defined above. This provides us rigorous and empirical views of the network evolution problem.

\section{Triad Position Profile}
\label{sec_triad_measure}
An important fraction of network dynamics locates in the process of influence evolution. A generic and effective measurement should be able to infer the influence evolution trend, and aid in predicting the potential prominence of a node in the future. We first introduce the current state of the art of influence measures and discuss their limitations. In addition, we validate the fundamental principles associated, and introduce our framework - {\bf triad position profile} which optimizes trade-offs between preferential attachment and triadic closure. Finally, based on experiments we unveil the interactions between the process of influence evolution and the process of link formation, which are well reflected in our framework.

\subsection{Current State of The Art}
\label{sec_current_stateofart}
The influence analysis in social networks has been a perennial topic of academic research. Typically these include influence maximization \cite{acm:influencemaximize0} \cite{acm:influencemaximize1}, influence selection and quantification \cite{acm:influenceselection} \cite{acm:goyal}, and influence validation \cite{acm:feedback}.

For influence maximization problem, there are quite a few influence diffusion models proposed, such as {\it linear threshold model} and {\it weighted cascade model} in the work of \cite{acm:influencemaximize0}. Many algorithms are designed to maximize the influence in these diffusion models, such as {\it DegreeDiscount} \cite{acm:influencemaximize1} and ``{\it CELF}'' \cite{acm:influencemaximize2}. At the same time many centrality measures are proposed for identifying influential nodes in a network, such as {\it degree centrality}, Pagerank \cite{acm:pagerank}, Betweenness \cite{acm:betweenness}, and Closeness \cite{acm:closeness}. In addition, Goyal et al. \cite{acm:goyal} studied the problem of learning influence probability of node from a log of user actions.

\subsubsection{Limitations of Current Methods} Although these methods are proved to be effective in influence quantification and measurement, they are inherently lack of predictability. First, most of these measures assign a value to each node, which leads to the loss of information; second, much research has focused on describing the influence at current time - that is, the consequence of influence evolution. This does not assure their predictability of future prominence. To summarize, even though these existing measures of influence degree are good at evaluating consequences of evolution, they have limitations in describing the future of influence evolution.

\subsubsection{A Case of Local Sub-structure}
Social influence is a well accepted phenomenon in social networks. We posit influence of a node, as well as capacity of a node being influenced, is a function of its neighborhood. Thus the future prominence of a node may be a function of the sub-structure surrounding the node at time $t$. We have several canonical examples to support this proposition. First, based on the {\it PageRank} heuristic: importance of a node is indicated by the number of connections or links to that node; second, Burt \cite{acm:structuralhole} proposed the concept of {\it structural hole}: a node's success often depends on their access to local bridges. Both of these examples imply that the position of node within a social network is important. This leads us to investigate the value of a node's position within the local sub-structures and the impact on its future prominence, which inspires the development of our framework.

\subsection{Preferential Attachment and Triadic Closure}
\label{sec_balance_theory}
\begin{figure}[t]
	\centering
		\includegraphics[width=2in]{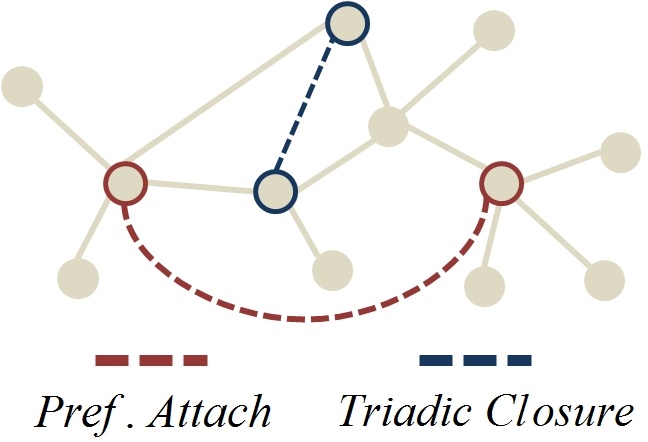}
\caption{Preferential Attachment vs. Triadic Closure. Based on the principle of {preferential attachment} two red nodes are most likely to be connected in future; while the {triadic closure} principle suggests that link between two blue nodes.}
  	\label{fig_pa_tc}
\end{figure}

Despite the well known macroscopic scaling in social networks, such as {\it power-law} degree distribution \cite{acm:preferential}, {\it attachment kernel} \cite{acm:attachmentkernel} and {\it clustering coefficient} as function of node degree \cite{acm:preferential}, it is undecided whether there is a common mechanism underlying these macroscopic laws \cite{acm:triad1} \cite{acm:triad2}. With the evidence that the {\it preferential attachment} process \cite{acm:preferential} is just one dimension of network evolution, much recent research has extended the {\it preferential attachment} principle by local sub-structure evolution rules \cite{acm:newman2001} \cite{acm:li2010}. Li et al. \cite{acm:li2010} and Jin et al. \cite{acm:newman2001} proposed that an individual's link formation significantly relies on its neighbors. In the work of \cite{acm:triadclosure}, Granovetter proposed that a ``forbidden'' triad (left in Figure~\ref{fig_tpp_example}) is most unlikely to occur in social networks, which means that the probability of a new link to close ``forbidden'' triad is higher than the probability of link between two randomly selected nodes. The principle of {\it triadic closure} is demonstrated to be relevant for social network evolution in many works \cite{acm:newman2001} \cite{acm:li2010} \cite{acm:transitivity} \cite{acm:triad1}. Obviously these two principles propose two distinct mechanisms of network evolution and none of them can act as a single origin of network evolution. In preferential attachment new links are made preferentially to high degree nodes while in triadic closure new links are generated to close ``forbidden'' triad (Figure~\ref{fig_pa_tc}). We are interested to know whether there is an effective combination of these two principles.

The principles of {\it preferential attachment} and {\it triadic closure} have been empirically demonstrated to be relevant (not as a single origin) for macroscopic scaling laws in the work of \cite{acm:jure2008} \cite{acm:validation1} \cite{acm:validation2} \cite{acm:validation3}, expressly or implicitly. As the fact that these principles are underlying the social network macroscopic scaling laws, we are interested to know whether these principles are valid to answer the microscopic problems in social network dynamics, such as {\it prominence prediction} and {\it link prediction}. Our work is different from the work of \cite{acm:jure2008} and \cite{acm:tang2012}, Leskovec et al. \cite{acm:jure2008} employ {\it triadic closure} to reproduce the observed macroscopic laws of social networks and Lou et al. \cite{acm:tang2012} investigated how a reciprocal link is developed and how relationships develop into triadic closure.

\subsubsection{Triadic Closure Effect on Network Evolution}
\begin{figure}[t]
	\centering
		\includegraphics[width=2in]{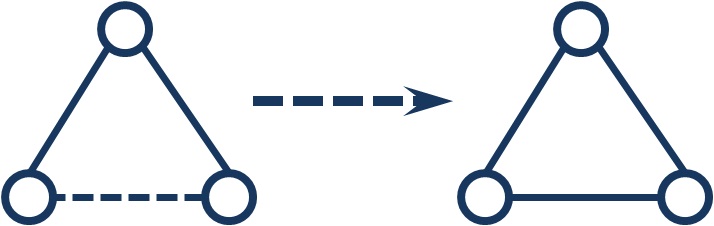}
\caption{Triadic Closure}
  	\label{fig_tpp_example}
\end{figure}
\begin{figure}[t]
	\centering
	\subfloat[Before Evolution]{\includegraphics[width=1.39in]{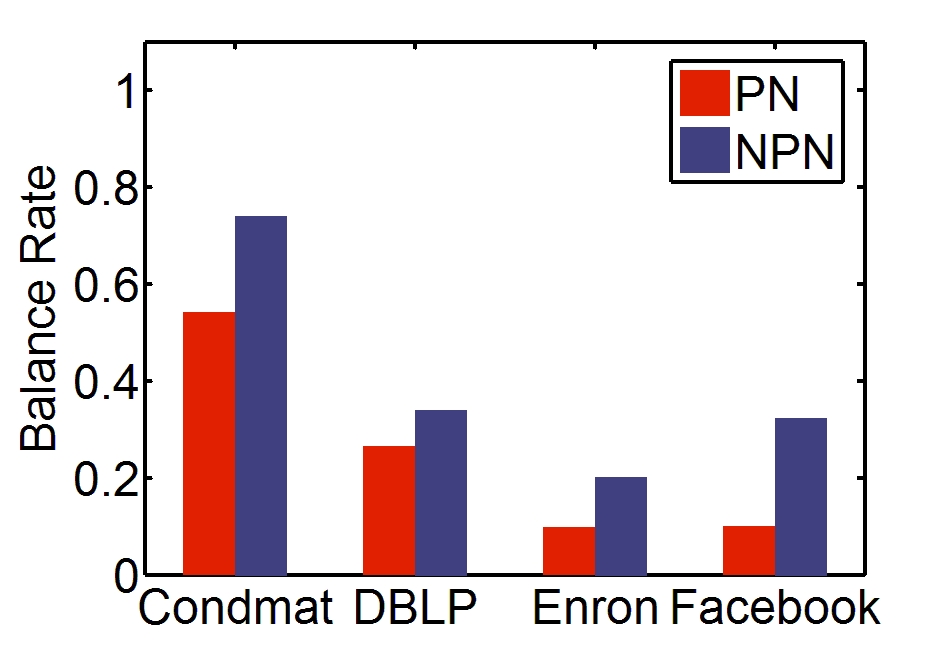}}\quad
	\subfloat[After Evolution]{\includegraphics[width=1.39in]{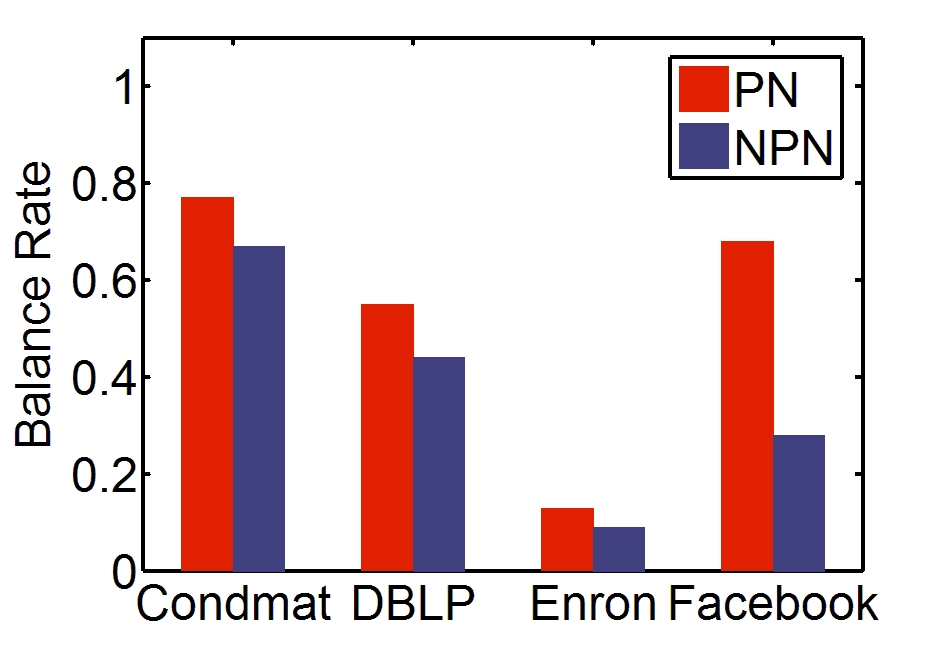}}\\
	\caption{Structural Balance Rate}
\label{fig_balance_rate}
\end{figure}

The effect of {\it preferential attachment} on the influence evolution is obvious and evident. Here we explore the effect of {\it triadic closure} on the influence evolution. The quantity of {\it triadic closure} (or structural balance) is usually defined as below \cite{acm:newman2001}:
\begin{equation}
\text{balance rate} = \frac{3 \times \text{number of closed triads}}{ \text{number of connected triads}} 
\end{equation}
where connected triad is the left triad in Figure~\ref{fig_tpp_example} and closed triad is the right triad in Figure~\ref{fig_tpp_example} respectively. By studying the sub-networks among prominent or non-prominent nodes, we observe that initially the future prominent nodes sub-network has a lower balance rate than the future non-prominent sub network, while after long enough evolution the prominent sub-network forms a more balanced structure (Figure~\ref{fig_balance_rate}). There are several implications:
\begin{compactenum}
\item There exists connections between the {\it triadic closure} and the prominence evolution. In addition, as discussed above, new links are more likely to form between nodes located in an imbalanced sub-network;
\item The initial sub-network where future prominent nodes are located is more imbalanced than that of future non-prominent nodes, position of node can be indicative of its future prominence.
\end{compactenum}
To some extent this implies the effect of {\it triadic closure} on both the influence evolution and the link formation mechanism.
\begin{figure}[t]
	\centering
	\subfloat[]{\includegraphics[height=0.8in]{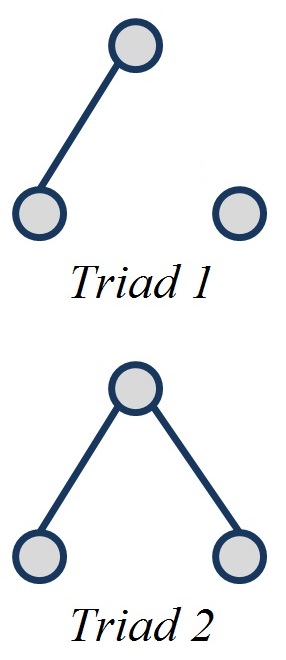}}\quad
	\subfloat[]{\includegraphics[height=0.8in]{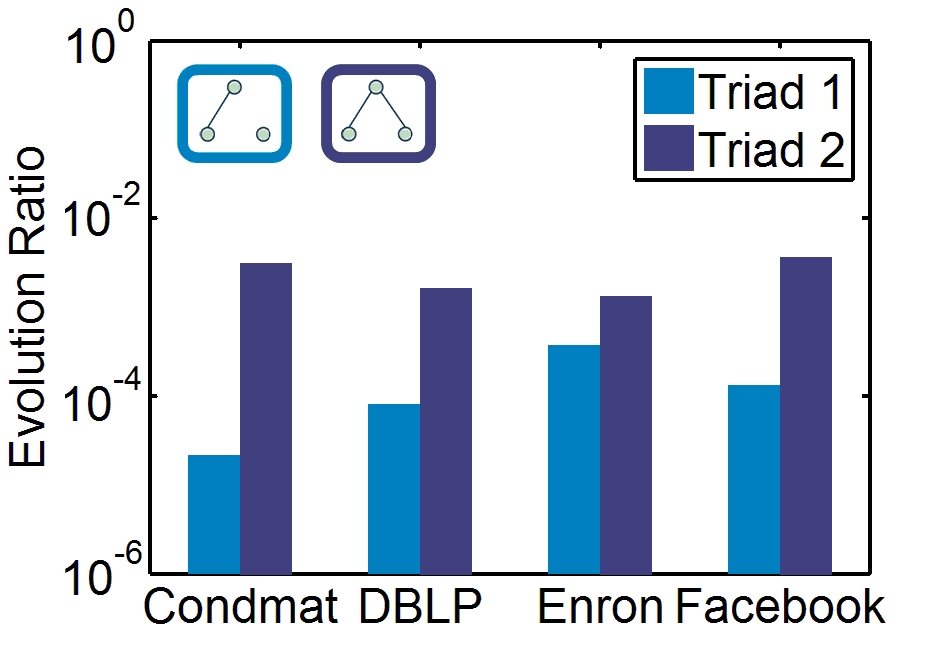}}\quad
	\subfloat[]{\includegraphics[height=0.8in]{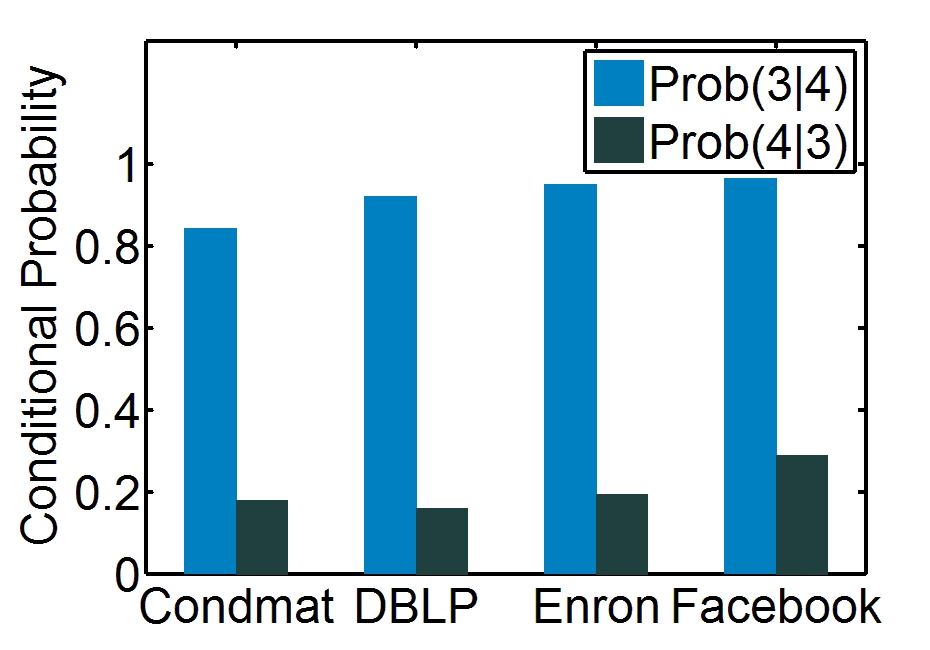}}\\
	\caption{Structural Balance Statistics}
\label{fig_triad_rate}
\end{figure}

As suggested in the principle of {\it triadic closure}, a ``forbidden'' triad is more likely to evolve into a closed triad. For further validation, we provide the evolution ratio of two types of triads in Figure~\ref{fig_triad_rate} (a). We can see that the ``forbidden'' triad ({\it triad 2}) has much higher probability to be a closure triad than the disconnected sub-structure {\it triad 1}. This implies, nodes in different triads have different probabilities to develop prominence and new links. This leads us to an important conclusion: the positions of nodes in sub-structures determine their future orbits in both essential evolution elements: the influence evolution and the link formation. This observation leads us to develop our method called, the Triad Position Profile, discussed in the next sub-section.

\subsection{Triad Position Profile}
\label{sec_influence_measure}
\begin{figure}[t]
\centering
	\subfloat[]{\includegraphics[height=1in]{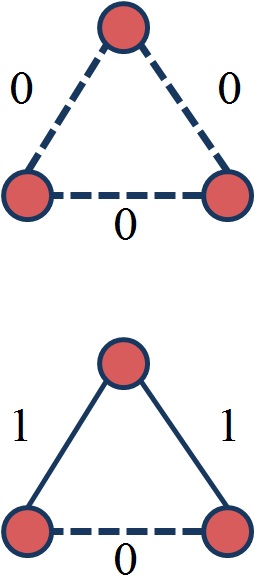}}\quad
	\subfloat[]{\includegraphics[height=1in]{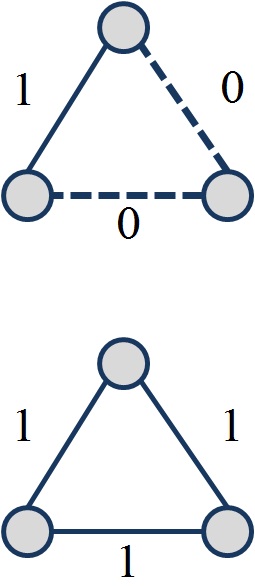}}\quad
	\subfloat[]{\includegraphics[height=0.8in]{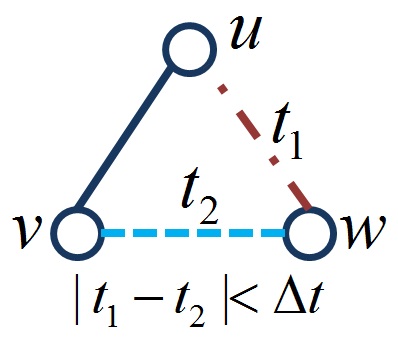}}\\	
	\caption{Triads and User Influential Probability.}
\label{fig_triad_example}
\end{figure}

Motivated by the above analysis, we start our investigations from the principle of {\it triadic closure}. Based upon the principle of {\it triadic closure}, an individual will try to close a ``forbidden'' triad that it has, for example in Figure~\ref{fig_tpp_example} a ``forbidden'' triad is likely to evolve as a closed triad. See examples of all possible triads in Figure~\ref{fig_triad_example}(a) and Figure~\ref{fig_triad_example}(b). The number labeled on the edge describes whether two nodes have relation, for instance `1' can state that two actors are friends while `0' means they are non-friends.

\begin{figure}[t]
\centering
		\includegraphics[width=2.3in]{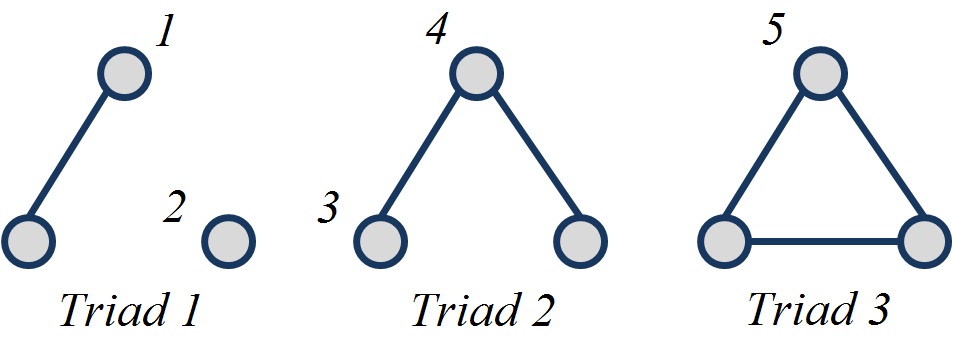}
\caption{Triad Position Profile}
  	\label{fig_tpp_example1}
\end{figure}

Such kind of triad evolution has very nice characteristics, firstly it leads to the formation of the link, and additionally it also increases the influence of node. Thus, different positions of a node in corresponding triads can be indicative of influence and prominence, as well as link formation analysis. This satisfies our proposition that the influence evolution and link formation are highly intertwined. As we have discussed above, the position of node within substructures could provide us insights into the principles underlying the network evolution. To that end, in Figure~\ref{fig_tpp_example1} we enumerate all possible five positions in the triad sub-structures for further study. We are interested to know that the consequences of preferential attachment and triad closure on the network evolution; second, we want to validate our proposition made in the above section and seek a solution which optimizes trade-offs between two distinct principles.

Based on our discussions above, we introduce our framework-{\it Triad Position Profile} for the influence evolution analysis. Formally, the {\it Triad Position Profile} is defined as follows:
\begin{mydef} {\bf Triad Position Profile}
{\it Triad Position Profile} for a node $v$, written as $\text{TPP} (v)$, is a vector describing the occurrence frequencies of node $v$ in five different positions in three isomorphic substructures.
\end{mydef}
In order to analyze the generality and effectiveness of existing influence measures and our method, we design an experiment to identify their correlation with node latent prominence. The evidence that our framework combining two principles well will be provided later.

{\bf Experimental Setup} For a time-varying network at time $t$ $G_{t}$, we extract the set of nodes whose arriving time is $t$ and then compute their influence measures based on the topology of $G_{t}$. At time $t+\Delta T$ for the network $G_{t+\Delta T}$ we classify the set of nodes into prominent set ${\bf PN}_{t+\Delta T}$ and non-prominent set ${\bf NPN}_{t+\Delta T}$ based on the topology of $G_{t + \Delta T}$. In order to demonstrate the discrimination of two principles, $\Delta T$ is selected large enough for the node influence evolution. As we know when $\Delta T$ is small the prominence prediction problem will be easy. Here we extract new arriving nodes as our prediction candidates, because existing nodes are well evolved and much easier to predict. In this way we can compare the correlations between these metrics and node's latent prominence quantitatively, we show the $p$-value associated with each feature and their corresponding significance level in Table~\ref{tab_significant}.
\begin{table}[t]
\caption{Significance of Features}
\centering
\resizebox{3.1in}{!}{
\begin{tabular}{|l|l|l|l|l|l|l|l|} \hline 
  {\bf Features} & $p$-value &significance level\\ \hline \hline
  Degree & 0.0583 & * \\ \hline
  Clustering Coefficient & 0.5053 &  \\ \hline
  Closeness Centrality & 0.7936 &  \\ \hline
  Betweenness Centrality & 0.0937 & *  \\ \hline
  PageRank & 0.1423 &  \\ \hline
  User Influential Probability & 0.2209 &  \\ \hline
  TPP Position 1 & 0.7388 &  \\ \hline
  TPP Position 2 & 0.0385 & **  \\ \hline
  TPP Position 3 & $1.059e^{-3}$ & *** \\ \hline
  TPP Position 4 & $1.55e^{-4}$ & **** \\ \hline
  TPP Position 5 & 0.31080 &   \\ \hline
\end{tabular}
}
\\
{
\begin{flushleft}
\scriptsize *: p $<$ 0.1; **: p $<$ 0.05; ***: p $<$ 0.01, ****: p $<$ 0.001.
\end{flushleft}
}
\label{tab_significant}
\end{table}

We observe (see Table~\ref{tab_significant}) that the centrality measures are not performing well in describing a node's future prominence except degree centrality and betweenness (1 sigma), while several {\it TPP} positions are significantly better in describing a node's latent prominence. For the {\it user influential probability} measure, the historic information of influence probability does not give a promising partition of ${\bf PN}$ and ${\bf NPN}$ (due to the lack of outside action log we calculate {\it user influential probability} as shown in Figure~\ref{fig_triad_example}(c), a link construction is considered as an action). We note that, in the experiment the sets of ${\bf PN}_{t+\Delta T}$ and ${\bf NPN}_{t+\Delta T}$ are labeled by the {\it degree centrality}, however we notice that {\it degree centrality} metric does not have a very significant correlation with node's future prominence. This implies that the {\it preferential attachment} is not the only dimension in the social network evolution as stated in \cite{acm:newman2001} \cite{acm:li2010} \cite{acm:jure2008}. While for the {\it TPP} positions, we have several observations: 1) we unfold that different {\it TPP} positions have different ability in describing node's future prominence; 2) three of {\it TPP} positions are much better than centrality measures. These observations hold for other datasets used in our work.

To {\bf summarize}, {{even though the centrality measures of influence degree are proved to be good at influence quantification, they are inherently not powerful enough to depict the node's future prominence.}} Additionally we can observe that the triad position profile combines two principles. Triad position $1$ and $4$ reflect the effect of preferential attachment, while triad position $3$ follows the triadic closure principle. This confirms our propositions made above and the effectiveness of our framework will be further validated.

As {\it triadic closure} principle suggests, for the unclosed triad ({\it triad $2$}) new links are formed between nodes in position $3$, however we have observed that nodes in position $4$ is more likely to be prominent in future. One possible reason underlying such phenomenon is the {\it preferential attachment} principle, nodes in position $4$ have higher attractiveness of links. However in Table~\ref{tab_significant} we observe that {\it degree centrality} does not have a comparable significance as the position $4$, this suggests that the {\it preferential attachment} principle is not the only mechanism underlying this.

To further study this effect, we calculated the conditional probability of position $3$ and position $4$, $Prob(3|4)$ states the probability that a node shows up in position $3$ given the condition that it is located in position $4$; $Prob(4|3)$ is the probability that a node is located in position $4$ given the condition that it is also in position $3$.  We can see in Figure~\ref{fig_triad_rate}(c) that nodes in position $4$ have extremely high probability to be located in position $3$ (close to 1.0), while nodes in position $3$ have less than $0.3$ probability to occur in position $4$. This means, nodes in position $4$ are affected by both mechanisms of {\it preferential attachment} and {\it triadic closure}, while nodes in position $3$ are mainly influenced by the {\it triadic closure} principle. This explains why position $4$ has higher significance level than position $3$, and further confirms that the {\it triadic closure} principle is more significant than the {\it preferential attachment} in social networks evolution. Also this implies an {\bf important} characteristic of the {\it TPP} method, the position profile combines two well know social principles (i.e. {\it preferential attachment} and {\it triadic closure}).
\subsection{Influence Evolution and Link Formation}
\label{sec_influence_event}
\begin{figure}[t]
	\centering
		\includegraphics[width=2.9in]{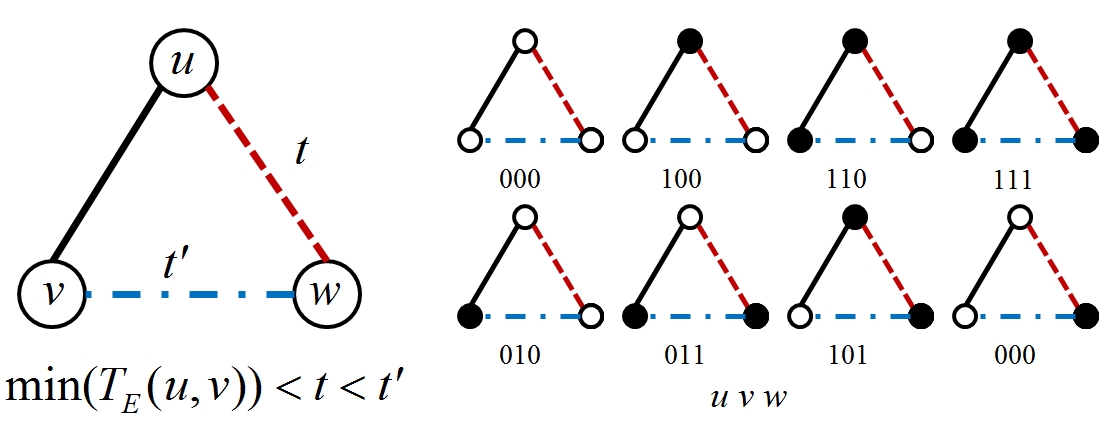}
\caption{Link Influence Events}
  	\label{fig_triad_event}
\end{figure}

As we conjectured in Section~\ref{sec_intro}, the influence evolution and the link formation are intertwined, and here we provide a detailed investigation into this from the perspective of influence events. Goyal et al. \cite{acm:goyal} proposed the concept of user influential probability that captures the influence degrees from the historic log of user actions. However such user actions are not always available in networks, here we define an action called {\it link action} between two nodes $u$ and $v$ as follows:
\begin{mydef}
For a given node $u$ in the time-varying network $G_{t} = (V, E, T_{V},T_{E})$, $u$ is said to have a {\bf link action} on node $w$ at time $t$ if $(u, w) \in E$ and $t \in T_{E} (u, w)$.
\end{mydef}
Additionally we provide the definition of the {\it link influence} of node $u$ on its neighbor $v$ as follows:
\begin{mydef}
A node $u$ is said to have a {\bf link influence} on its neighbor $v$ {\it iff}: 1) there is a link action of node $u$ with another node $w$ at time $t$; 2) there exists a link action of node $v$ with node $w$ at time $t'$; 3) $min(T_{E}(u,v)) < t < t'$ and $t' - t < \sigma$
\end{mydef}
The $\sigma$ is the average action delay between two nodes $u$ and $v$. An example of {\it link influence} is presented in Figure~\ref{fig_triad_event} (left).

In our work we divide nodes into two groups (prominent nodes and non prominent nodes), in this section we further study the connection between the node's prominence and the link formation. Thus in Figure~\ref{fig_triad_event} we partition the {\it link influence} event into $8$ categories based on nodes' prominence. The three digits represent the prominence status of the three nodes, $u$, $v$, and $w$, `1' indicates prominent node and `0' indicates non prominent node. In Table~\ref{tab_triad_event_stat} we provide the distribution of several patterns, and we observe that: 1) $|1XX| > |0XX|$ and $|X1X| > |X0X|$, this means prominent nodes have much higher probability to have {\it link influence} on their neighbors, and it also validates the principle of {\it preferential attachment}; 2) additionally $|XX0| > |XX1|$, non-prominent nodes play an important role to transfer {\it link influence}; 3) $|11X| > |00X|$, this states that {\it link influence} is more likely to happen between prominent nodes; 4) $|10X| \approx |01X|$, if {\it link influence} occurs among prominent nodes and non-prominent nodes, then prominent nodes and non-prominent nodes have the same chance to initiate the influence. To {\bf summarize}, this validates the intimate interactions between the influence evolution and the link formation. Thus we postulate to validate the effectiveness of our framework in these two microscopic problems.
\begin{table}[t]
\caption{\bf{Prominence Status vs. Link Influence Event}}
\label{tab_triad_event_stat}
\centering
\resizebox{3.2in}{!}{
\begin{tabular}{|r|r|r|r|r|r|r|r|r|r|r|} \hline
{\bf Patterns} & 1XX & 0XX & X1X & X0X & XX1 & XX0 & 11X & 00X & 10X & 01X \\ \hline \hline
Condmat &1530 &365 & 1513 & 382 & 95 & 1800 & 1316 & 168 & 214 & 197\\ \hline
DBLP &1377 &438 & 1329 & 486 & 15 & 1800 & 681 & 498 & 369 & 267\\ \hline
Enron &11769 &249 & 11787 & 231 & 187 & 11831 & 11549 & 11 & 220 & 238\\ \hline
Facebook &6203 &2775 & 6196 & 2782 & 10 & 8977 & 4794 & 1373 & 1409 & 1402\\ \hline
\end{tabular}
}
\end{table}
\section{Inferring Future Prominence}
\label{sec_infer_prominence}
In order to prove the correctness of our framework, we apply our approach in prominent prediction problem and compare with baseline methods. Note that we classified nodes as PN or NPN, thus making it a binary classification task. We first discuss the feature vector construction aspect.

\subsection{Feature Vector Engineering}
We first integrate the various measures capturing the notion of influence in to one feature vector. In addition to the different measures described in Table~\ref{tab_significant} (other than TPPs), we also include some measures introduced in Burt's work of \cite{acm:structuralhole}, such as {\it efficiency}, {\it constraint} and {\it hierarchy}. These features contribute to the feature vector for the {\bf Baseline} method. 

The five TPP positions census contributes to our {\bf TPP} method for prediction. In addition,  we developed a method based on triad substructure influence census ({\bf TPP+}), as follows. 

We first compute the {\it link influence probability} of a node $u$, which can be expressed as 
\begin{equation}{LIP} (u) = \frac{|\text{link influence} (u)|}{| \text{link action} (u) |} \end{equation}

In Figure~\ref{fig_triad_event} we can see that a node $u$ with high {\it LIP} is more likely to attract links for its neighbors. Our heuristic is, if a node has large number of connections with high {\it LIP} nodes then it has higher probability to be prominent in future. Based on this heuristic we design two features to describe a node's prominence trend:
\begin{align*}
&prominence\_prob (v) = 1 - \prod (1 - \text{LIP} (u)), (u, v) \in E\\
&prominence\_index (v) = \sum \text{LIP} (u), (u, v) \in E
\end{align*}
Thus, TPP+ comprises of TPP, as well as $prominence\_prob$ and $prominence\_index$.  

The features for {\it Baseline} method, {\it TPP} and {\it TPP+}  are listed in Table~\ref{table_feature_list}. For all methods, we use Bagging with {\it Logistic Regression} as the supervised learning model. Our goal here is to evaluate the utility of additional information imputed by us in the feature vector versus the quality of a learning algorithm. We conjecture that the benefits of another learning algorithm may uniformly apply to the task, and provide improvements across the board. 
\begin{table}[t]
\caption{\bf{Features List}}
\label{table_feature_list}
\centering
\resizebox{2.3in}{!}{
\begin{tabular}{|l|l|l|l|} \hline
{\bf Features} & {\bf Baseline} & {\bf TPP} & {\bf TPP+} \\ \hline \hline
Degree & $\surd$ & & \\ \hline
Betweenness &  $\surd$ &  &\\ \hline
Closeness & $\surd$ & & \\ \hline
Clustering Coef. &  $\surd$ &  &\\ \hline
PageRank & $\surd$ & &\\ \hline
Efficiency &  $\surd$ & & \\ \hline
Hierarchy &  $\surd$ &  &\\ \hline
Constraint &  $\surd$ &  &\\ \hline
TPP & & $\surd$ &  $\surd$  \\ \hline
prominence\_prob  & &  & $\surd$  \\ \hline
prominence\_index & &  & $\surd$  \\ \hline
\end{tabular}
}
\end{table}

\subsection{Experimental Settings}
In our experiment we only allow methods to observe features of nodes in a short duration after nodes arriving, for example, for Condmat and DBLP we only use the first year data of new arriving nodes and for Enron and Facebook we only use the first month data of new arriving nodes. We classify the nodes in to PN and NPN using {\it degree centrality}.
\begin{figure*}[t]
	\centering
	\subfloat[Condmat-LM]{\includegraphics[width=1.45in]{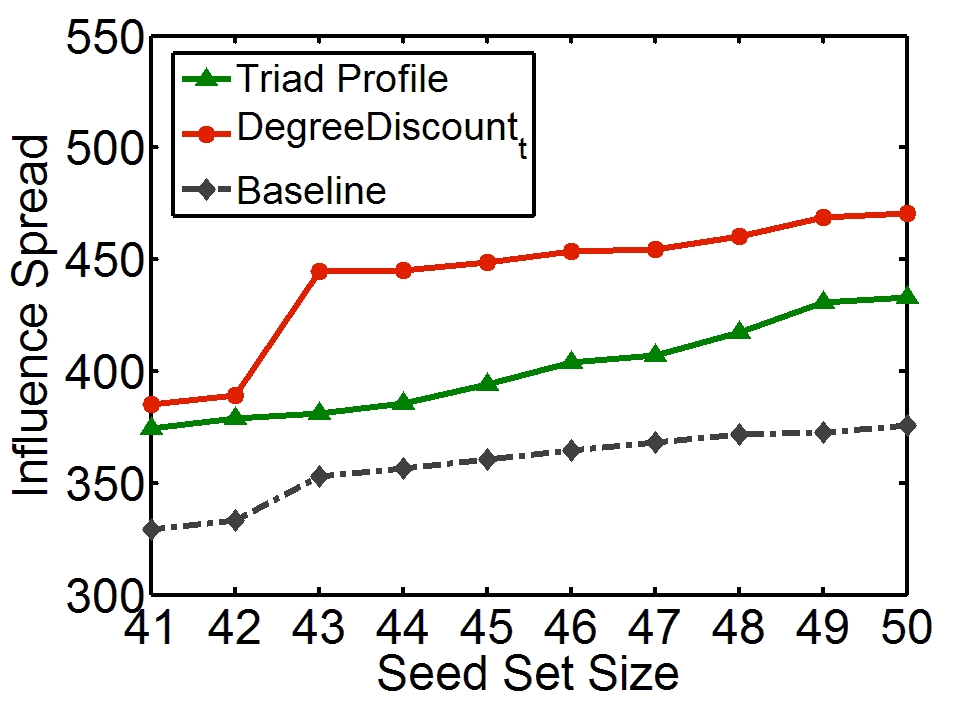}}\quad
	\subfloat[DBLP-LM]{\includegraphics[width=1.45in]{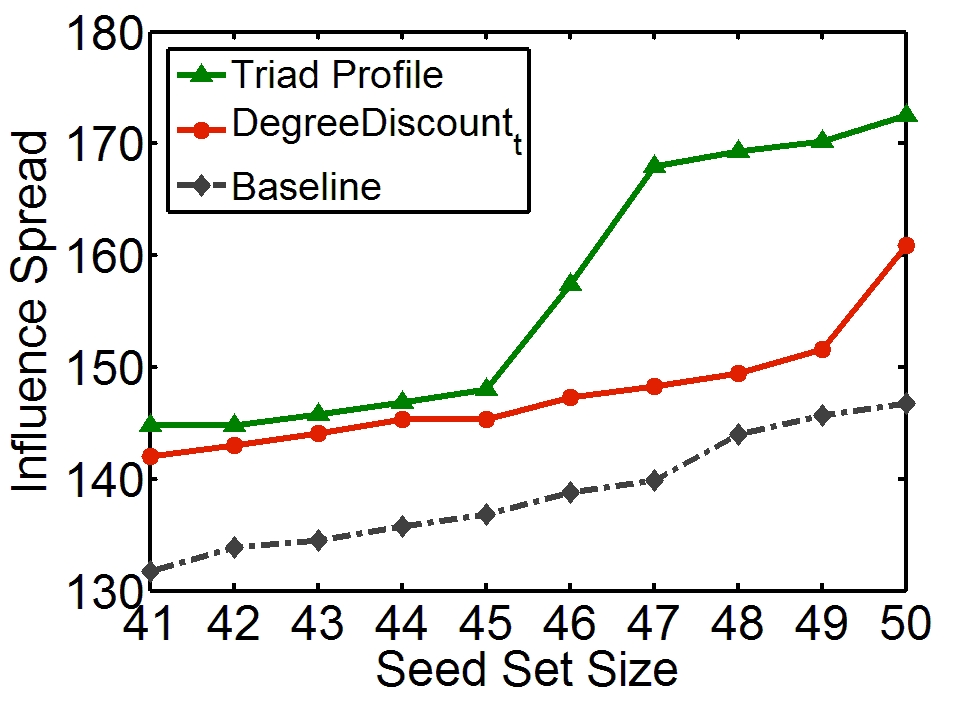}}\quad
        \subfloat[Enron-LM]{\includegraphics[width=1.45in]{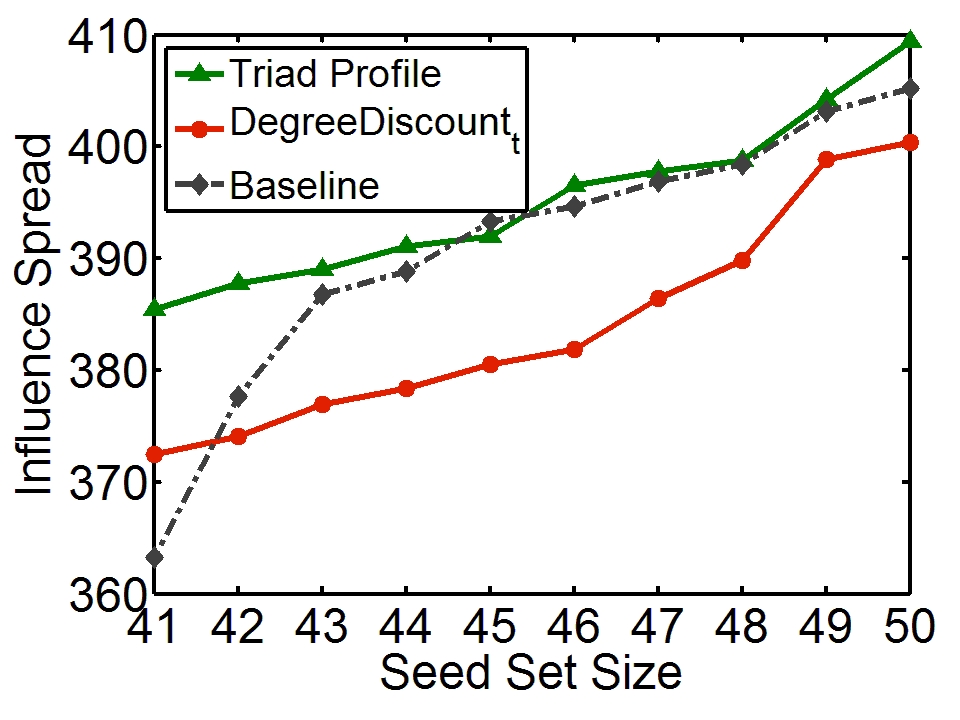}}\quad	
	\subfloat[Facebook-LM]{\includegraphics[width=1.45in]{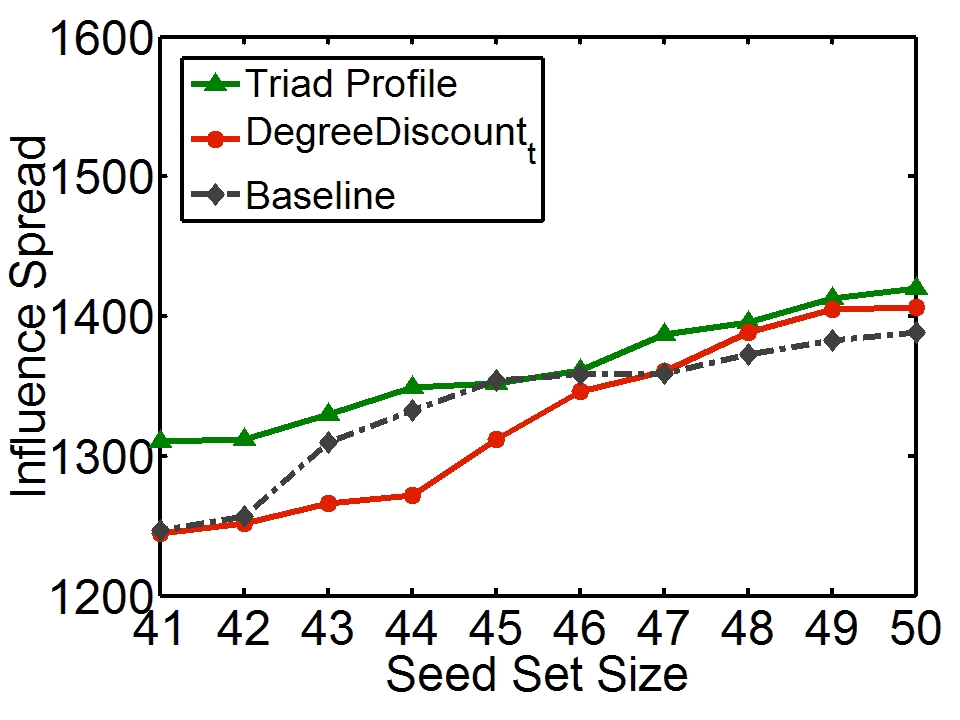}}\\
	\subfloat[Condmat-WM]{\includegraphics[width=1.45in]{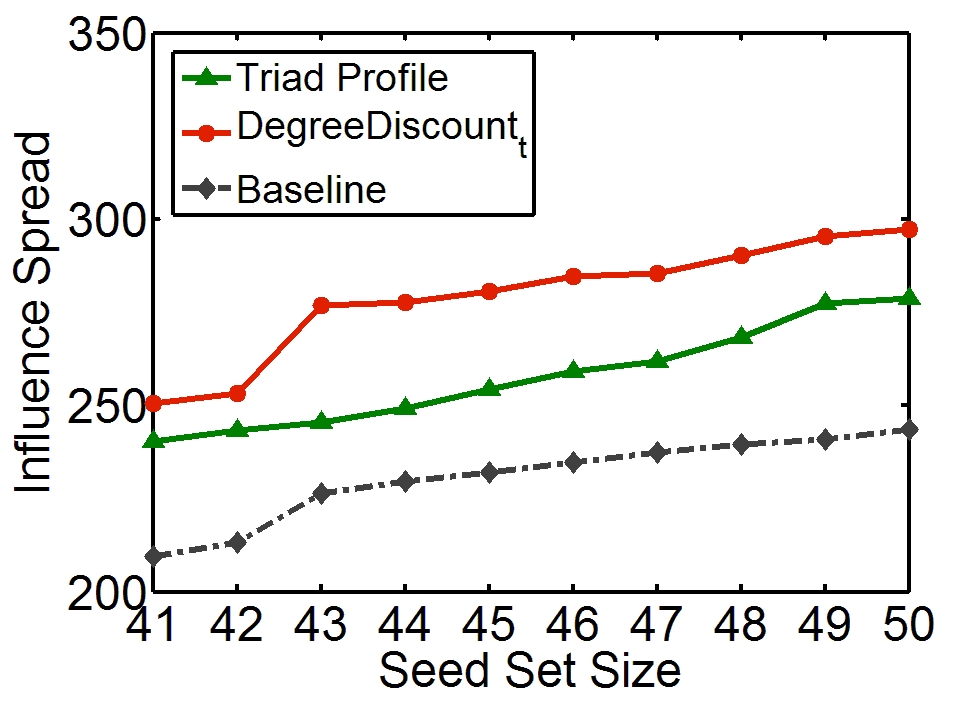}}\quad
	\subfloat[DBLP-WM]{\includegraphics[width=1.45in]{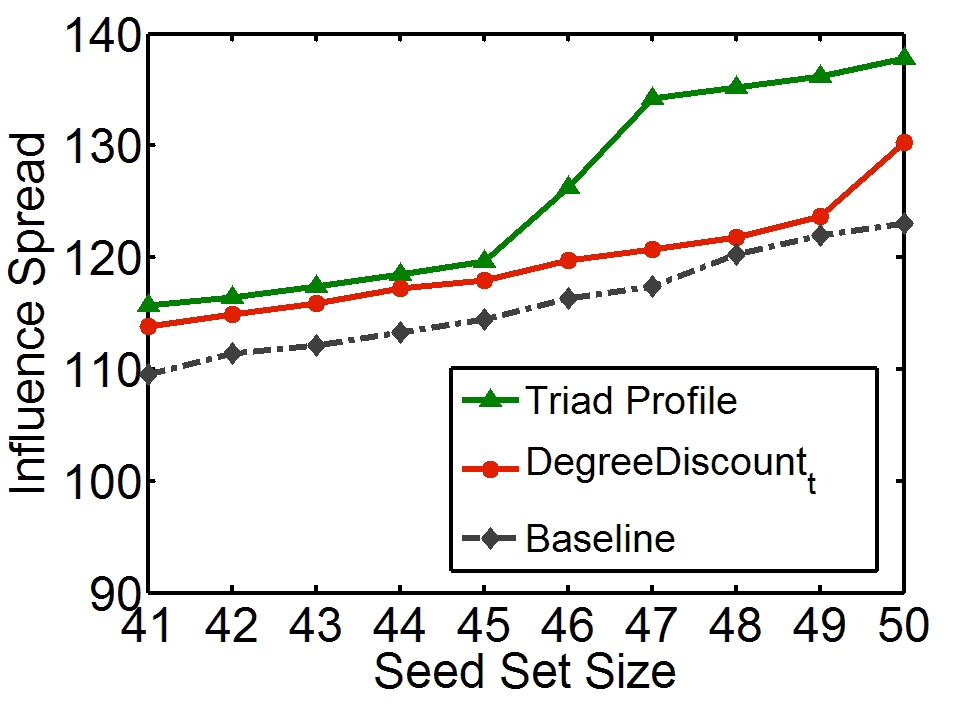}}\quad
        \subfloat[Enron-WM]{\includegraphics[width=1.45in]{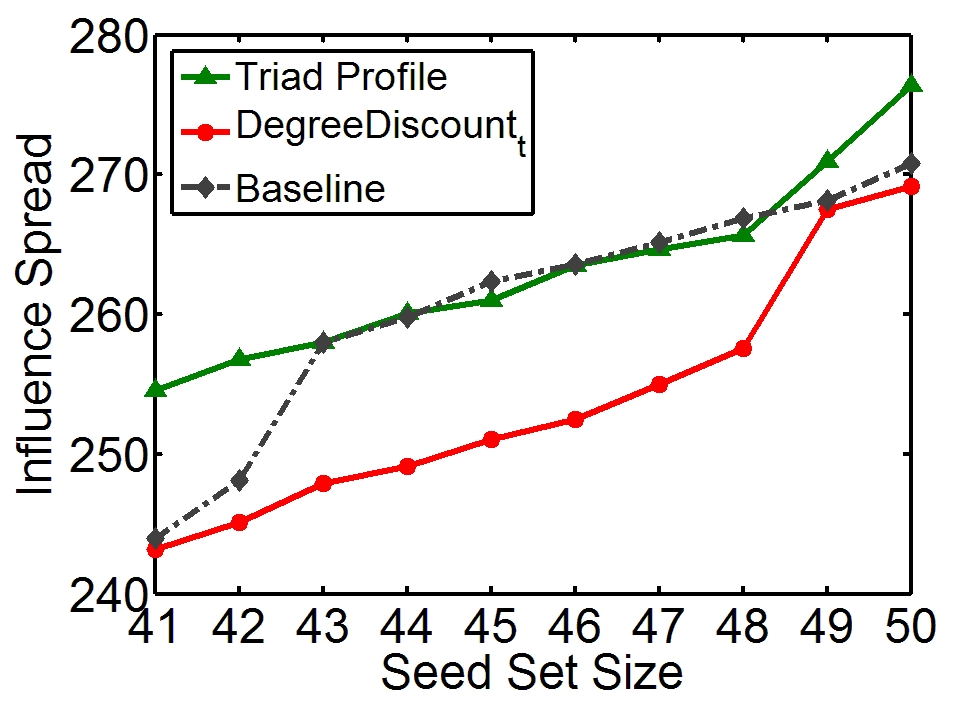}}\quad	
	\subfloat[Facebook-WM]{\includegraphics[width=1.45in]{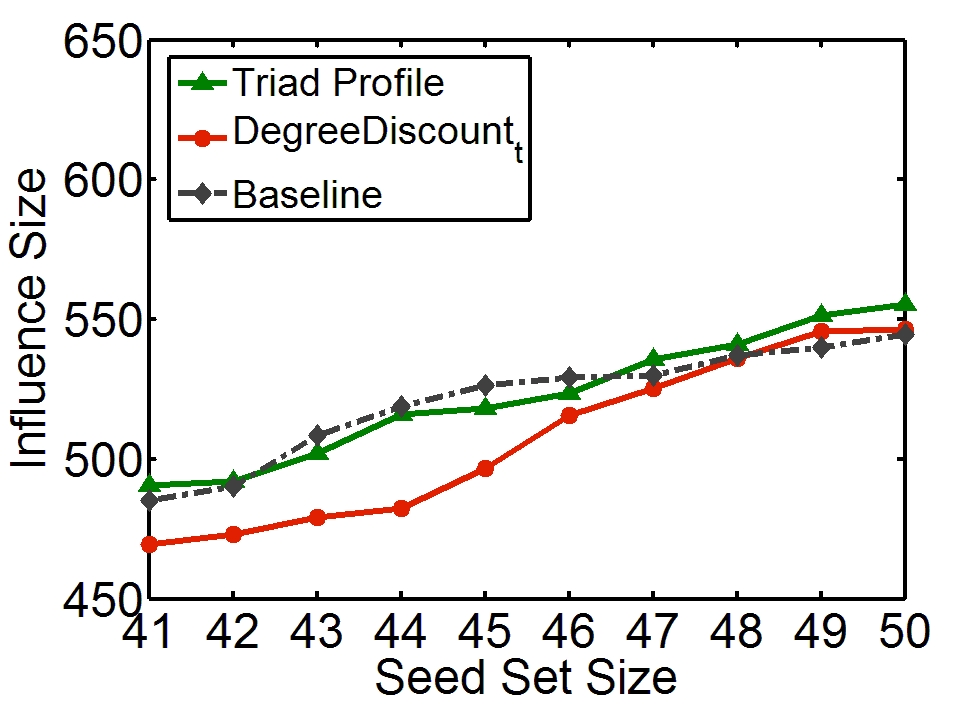}}\\
	\caption{Linear Threshold Model and Weighted Cascade Model (degree labeling). X-axis represents the seed set size and Y-axis is the influence spread size.}
\label{fig_degree_exp}
\end{figure*}

\subsection{Prominence Prediction}
\subsubsection{Classification Performance}
\begin{table}[t]
\caption{Performance Comparisons} 
\label{performance_metric1}
\centering 
\resizebox{3.3in}{!}{
\begin{tabular}{|c|p{1cm}|p{0.7cm}|p{1cm}|p{0.7cm}|p{1cm}|p{0.7cm}|p{1cm}|p{0.7cm}|} 
\hline
{} & \multicolumn{2}{c|}{Accuracy} & \multicolumn{2}{c|}{AUC} & \multicolumn{2}{c|}{AUPR} & \multicolumn{2}{c|}{Top@50}\\
\cline{2-9}
{Datasets} & {Baseline} & {TPP+} & {Baseline} & {TPP+} & {Baseline} & {TPP+} & {Baseline} & {TPP+}\\
\hline
Condmat & 0.880 & {\bf 0.893} & 0.853 & {\bf 0.858} & 0.712 & {\bf 0.717} & {\bf 50} & {\bf 50}  \\ \hline
DBLP & {\bf 0.912} & 0.898 & 0.724 & {\bf 0.817} & 0.189 & {\bf 0.220} & 7 & {\bf 10} \\ \hline
Enron & 0.156 & {\bf 0.856} & 0.704 & {\bf 0.711} & 0.506 & {\bf 0.519} & 38 & {\bf 42} \\ \hline
Facebook & 0.965 & {\bf 0.970} & {\bf 0.738} & 0.730 & 0.417 & {\bf 0.430} & {\bf 8}  & {\bf 8} \\ \hline
\end{tabular}
}
\end{table}

In Table~\ref{performance_metric1}, we provide an empirical comparison of learning performance. In our observation our approach {\it TPP}+ outperforms the {\it baseline} method in terms of {\it AUPR} and {\it Top@50}, and has better or comparable performance in terms of {\it AUC} and {\it Accuracy}. The {\it TPP}+ improves {\it AUPR} by 0.7\%-16.4\% and improves {\it Top@50} by 0\%-42.8\%. This confirms that our approach has generalized performance in different domains of datasets. The performance of {\it TPP} is provided in section~\ref{sec_generalization}, which is also better than the {\it baseline} method. We have several {\bf conclusions}: 1) the principle {\it preferential attachment} is just one dimension of mechanisms underlying the nodal influence evolution; 2) the trade-offs between {\it triadic closure} and {\it preferential attachment} are well balanced in {\it triad position profile} and then it achieves better performance in prediction task.

\subsubsection{Impact of Different Influence Models}
There are different centrality and influence measures for evaluating the prediction of a node's prominence, and it is not possible to enumerate performance across each of those dimensions. To resolve that and do a robust evaluation,  we used the influence propagation models for further validation.

In the prominence prediction problem, our task is to predict whether the set of nodes arriving at time $t$ (denoted as {\bf NAN} (new arriving nodes)) will become prominent at time $t + \Delta T$. The prominence predictors will rank the nodes in {\bf NAN} based on their likelihoods of being prominent in the future. When applied in influence maximization problem, a simple method is to extract the top $k$ ranked nodes (based on different metrics, i.e., degree) as the seed set. The seed set extracted from our approach is denoted as {\it triad profile} while for the {\it baseline} method we denote such a set as {\it baseline}. Comparing these two seeds sets' influence spread in future network $G_{t + \Delta T}$ will suggest which method is a better indicator of future prominence. Besides this, we also build a reference system for the validation of predictability. We employ the {\it DegreeDiscount} \cite{acm:influencemaximize1} (an efficient and scalable algorithm in influence maximization) to identify the top $k$ seed set from the set {\bf NAN} based on the topology of network $G_{t}$, this top $k$ seed set is denoted as $\text{DegreeDiscount}_{t}$. If the seed set extracted from a prominence predictor' results has better influence spread than $\text{DegreeDiscount}_{t}$, then we consider this predictor owns predictability in dynamic influence maximization.

As shown in Figure~\ref{fig_degree_exp}, our approach still outperforms the {\it baseline} method. In datasets DBLP, Enron and Facebook, our approach reveals its predictability in dynamic influence maximization, while in Condmat the $\Delta T$ is too short thus the benefit of predictability is not significant. To note that, our method is not designed for the purpose of influence maximization, the comparisons in influence propagation model are employed to provide an empirical and solid comparison between {\it baseline} method and {\it TPP}+.

\section{Link Prediction}
\label{sec_link_prediction}
As we discussed above, the influence evolution and the link formation/dissolution are intimately connected. In this section we demonstrate that our framework can also interpret the link formation process better than has heretofore been possible.

\subsection{Feature Vector Engineering}
\subsubsection{Triad Evolution Matrix Predictor}
\begin{figure}[t]
	\centering
	\subfloat[Triads]{\includegraphics[height=1.2in]{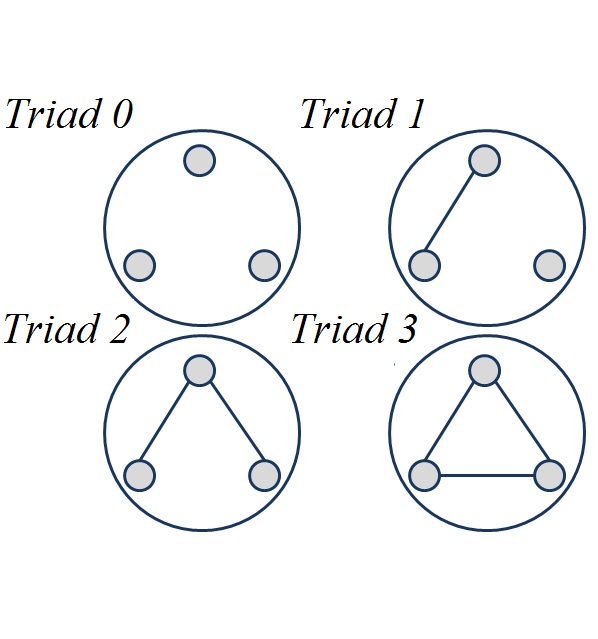}}\quad
	\subfloat[Triads Evolution]{\includegraphics[height=1.2in]{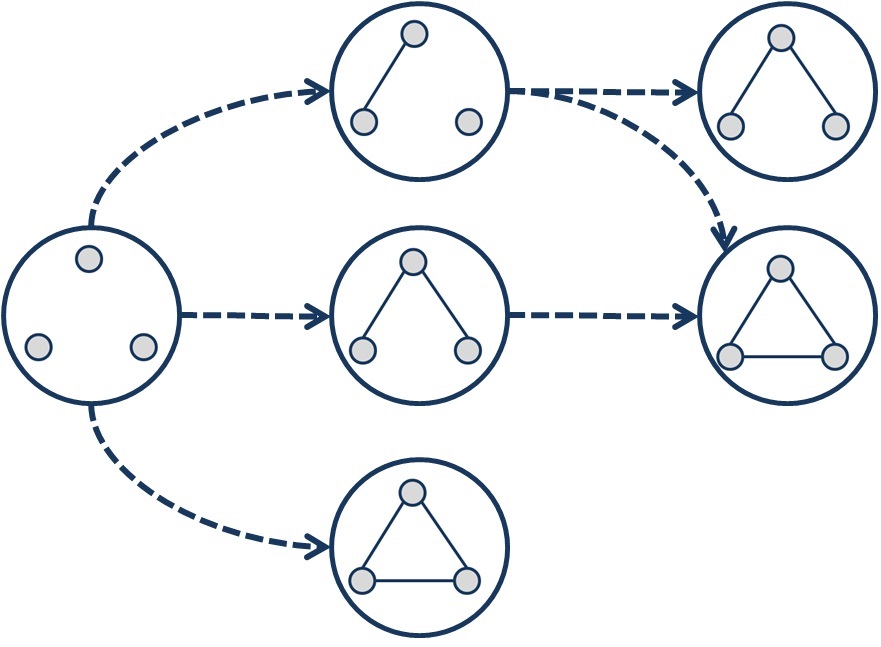}}\\
	\caption{Triads and Triads Evolution.}
\label{fig_triads}
\end{figure}

Motivated by our discussion in Section~\ref{sec_balance_theory} and Section~\ref{sec_influence_event}, here we introduce a method called {\it Triad Evolution Matrix (TEM)}, which are adapted from triad position profile to perform the link prediction task. All possible 3-subgraph in an undirected network is presented in Figure~\ref{fig_triads} (a), and their transition relationships are provided in Figure~\ref{fig_triads} (b). Based on our observations in Section~\ref{sec_balance_theory} different kinds of triads have different probabilities to be closure triad, this inspires us to perform the census of nodes pair's collocation in these triads and gain discernment in predicting new links.

\begin{figure}[t]
	\centering
	\subfloat[TEM]{\includegraphics[height=1.4in]{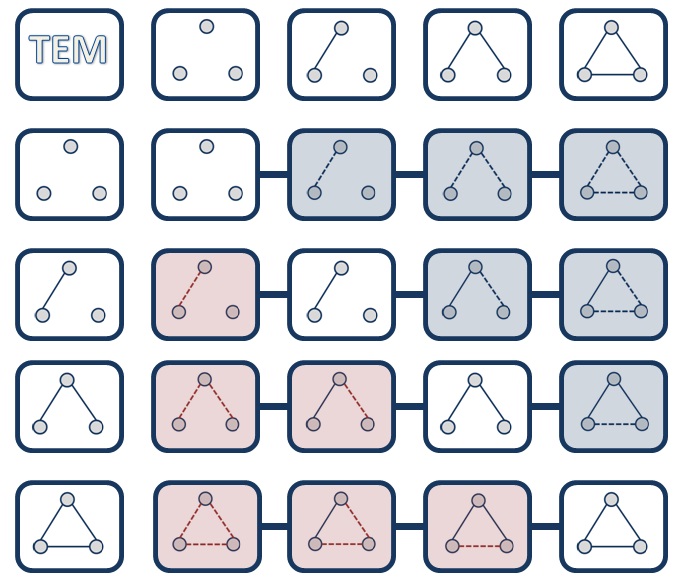}}\quad
	\subfloat[TCE]{\includegraphics[height=1.4in]{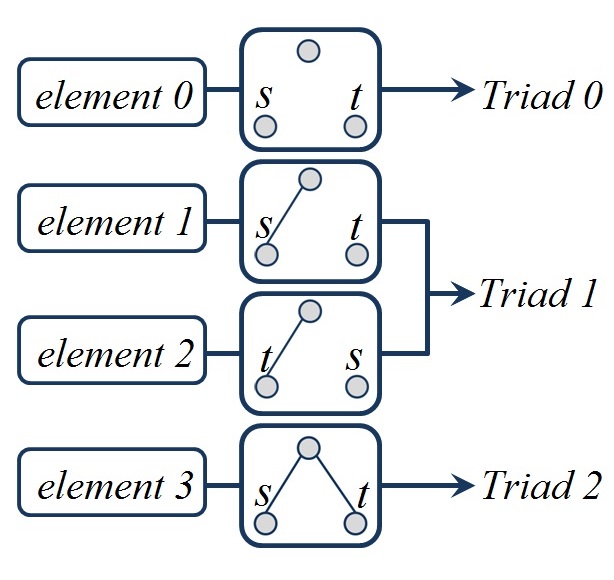}}\\
	\caption{Triads Evolution Matrix and Triad Collocation Elements.}
\label{fig_triads_evolution_matrix}
\end{figure}

The TEM is a matrix of size $n \times n$ ($n=4$), where n is the number of triads in undirected networks, and TEM$[i,j]$ represents the percentage of triad-$i$ at time $t$ evolve to triad-$j$ at time $t+1$ (Figure~\ref{fig_triads_evolution_matrix} (a)). This matrix can be computed trivially by counting triads in the network $G_{t}$ at time $t$ and then calculating elements by checking the network $G_{t+1}$ at time $t+1$. Additionally in the network $G$ there are four possible {\itshape triad collocation elements} (Figure~\ref{fig_triads_evolution_matrix} (b)) for two nodes $s$ and $t$ where $e(s,t) \notin G$. Thus for any two nodes $s$ and $t$ collocated in TCE$\_i$ ({\itshape triad collocation elements}) ($i \in \{0,1,2,3\}$), we can compute the likelihood of potential link between $s$ and $t$ as below:
\begin{equation}
\begin{split}
l ^{TCE\_i} = \begin{cases}
&(0,\frac{1}{3},\frac{2}{3},1) \cdot TEM[0,.] \text{ if } i=0 \\
&(0,0,\frac{1}{2},1) \cdot TEM[1,.] \text{ if } i=1,2 \\
&(0,0,0,1) \cdot TEM[2,.] \text{ if } i=3
\end{cases}
\end{split}
\end{equation}
In this way we get a likelihood vector for TCE,
\begin{align*}
 l=(l ^{TCE\_0}, l ^{TCE\_1}, l ^{TCE\_2}, l ^{TCE\_3})
\end{align*}

To note that here we are using the dot product of two vectors of size 4, the result will be a real number. And the case that two nodes $s$ and $t$ are collocated in TCE$\_1$ is equivalent to the case that they are collocated in TCE$\_2$ for undirected link prediction.

Thus for two nodes $s$ and $t$ we can calculate a probability vector based on the corresponding TCE vector,
\begin{align*}
\text{TCE}_{s,t} = (|TCE\_0_{s,t}|,|TCE\_1_{s,t}|,|TCE\_2_{s,t}|,|TCE\_3_{s,t}|)
\end{align*}
,where $|TCE\_i_{s,t}|$ states for how many TCE$\_i$ elements nodes $s$ and $t$ are collocated in. The corresponding probabilities vector $TEM\_prob (s,t)$ can be written as:
\begin{equation}
\begin{split}
TEM\_prob (s,t) = (TCE_{s,t}[0] \cdot l[0], TCE_{s,t}[1] \cdot l[1], \\
TCE_{s,t}[2] \cdot l[2], TCE_{s,t}[3] \cdot l[3])
\end{split}
\end{equation}
In this way this vector gives a multi-dimensional description of the link likelihood of a pair of nodes $s$ and $t$ based on the principle of {\it triadic closure}, and which also balances well with the {\it preferential attachment}.

\subsubsection{Link Influence Census}
As we discussed in Section~\ref{sec_infer_prominence}, we can calculate the {\it LIP} for a individual node using the census of {\it link influence}. Similarly, we can compute the {\it LIP} of a node $u$ on its neighbor $v$, as follows:
\begin{align*}
&\text{LIP} (u, v) = \frac{|\text{link influence} (u, v)|}{|\text{link action} (u,.)|}, (u, v) \in E
\end{align*}
Thus for a pair of nodes $v$ and $w$, we can calculate their link likelihood based on our observed $\text{\it link influence probability}$ information and most recent link actions within $\Delta t$ time. Firstly for the common neighbor node $u$ of $v$ and $w$, we need to calculate the probability of link between $v$ and $w$ due to the influence of $u$.
\begin{align*}
&p_{v,w}^{u,v} = 1- (1 - \text{LIP} (u, v))^{|{t' | t' \in T_{E}(u,w)} \wedge t' > t-\Delta t|}\\
&p_{v,w}^{u,w} = 1- (1 - \text{LIP} (u, w))^{|{t' | t' \in T_{E}(u,v)} \wedge t' > t-\Delta t|}\\
&p_{v,w}^{u} = max (p_{v,w}^{u,v}, p_{v,w}^{u,w})
\vspace{-0.3cm}
\end{align*}

In the above equations, $p_{v,w}^{u,v}$ represents the probability of link between $v$ and $w$ due to the influence probability of $u$ on $v$, while $p_{v,w}^{u,w}$ represents the probability of link between $v$ and $w$ due to the influence probability of $u$ on $w$. Trivially, we can use the maximum value of them to represent the probability of link between $v$ and $w$  due to the node $u$, denoted as $p_{v,w}^{u}$. And for a pair of nodes $v$ and $w$ there can exist many common neighbors, thus we define the probability of link between $v$ and $w$ due to the link influence effect as follows:
\begin{align}
\resizebox{0.9\hsize}{!}{$
prob(v,w) = 1 - \prod (1 - p_{v,w}^{u}), (u, v) \in E \wedge (u, w) \in E
$}
\label{eqn_lp}
\end{align}

\begin{table}[t]
\caption{\bf{Features List}}
\centering
\resizebox{2.5in}{!}{
\begin{tabular}{|l|l|l|l|l|l|l|l|} \hline
{\bf Features} & {\bf TEM-}&{\bf TEM} & {\bf TEM+} \\ \hline \hline
$TCE_{s,t}$ &$\surd$ & $\surd$ & $\surd$ \\ \hline
$TEM_{prob}(s,t)$ & & $\surd$ &  $\surd$ \\ \hline
Link Influence Prob. (Equation~\ref{eqn_lp}) & & &  $\surd$ \\ \hline
\end{tabular}
}
\label{table_feature_list1}
\end{table}
\begin{table}[t]
\caption{Performance Comparisons} 
\label{performance_metric}
\centering 
\resizebox{1\hsize}{!}{
\Large
\begin{tabular}{|c|c|c|c|c|c|c|c|c|c|c|c|c|} 
\hline
 & \multicolumn{6}{|c|}{Condmat} & \multicolumn{6}{|c|}{DBLP} \\
\cline{2-13}
{Methods} & {HPLP} & {TEM} & {TEM+} & {AA} & {CN} & {PA} & {HPLP} & {TEM} & {TEM+} & {AA} & {CN} & {PA}\\ \hline
AUPR & 0.010 & 0.015 & {\bf 0.015} & 0.008 & 0.009 & 0.006 & 0.014 & 0.020 & {\bf 0.031} & 0.009 & 0.013 & 0.009  \\ \hline
 & \multicolumn{6}{|c|}{Enron} & \multicolumn{6}{|c|}{Facebook}\\
\cline{2-13}
{Methods} & {HPLP} & {TEM} & {TEM+} & {AA} & {CN} & {PA} & {HPLP} & {TEM} & {TEM+} & {AA} & {CN} & {PA}\\ \hline
AUPR & 0.160 & {\bf0.243} & 0.211 & 0.050 & 0.153 & 0.036 & 0.042 & 0.050 & {\bf 0.050} & 0.023 & 0.036 & 0.010  \\ \hline
\end{tabular}
}
\label{tab_lp_results}
\end{table}

\begin{figure*}[t]
	\centering
	\subfloat[Condmat-ROC]{\includegraphics[height=1.15in]{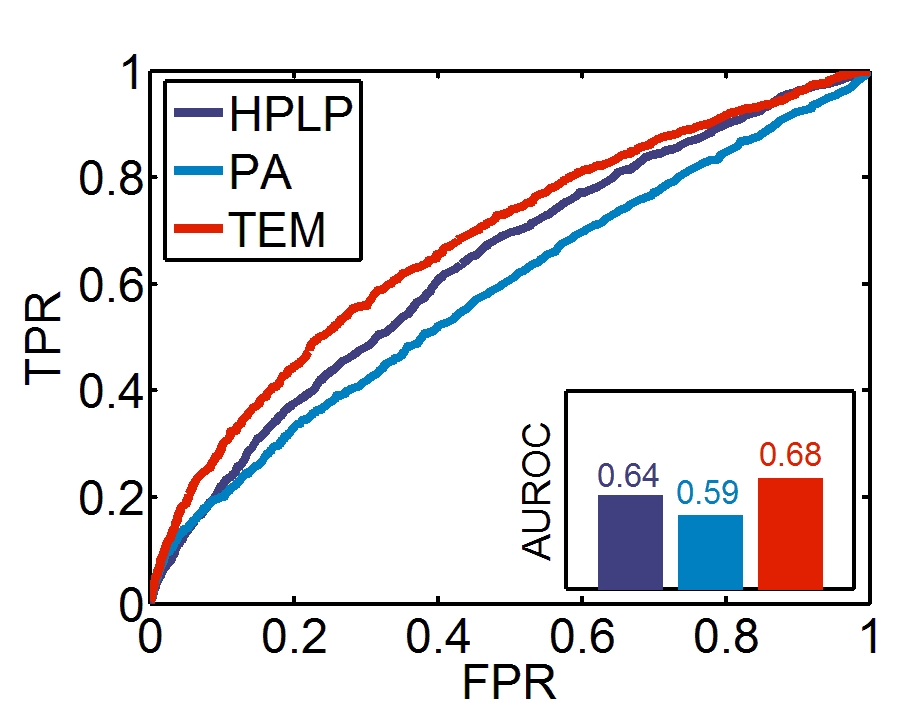}}\quad
	\subfloat[DBLP-ROC]{\includegraphics[height=1.15in]{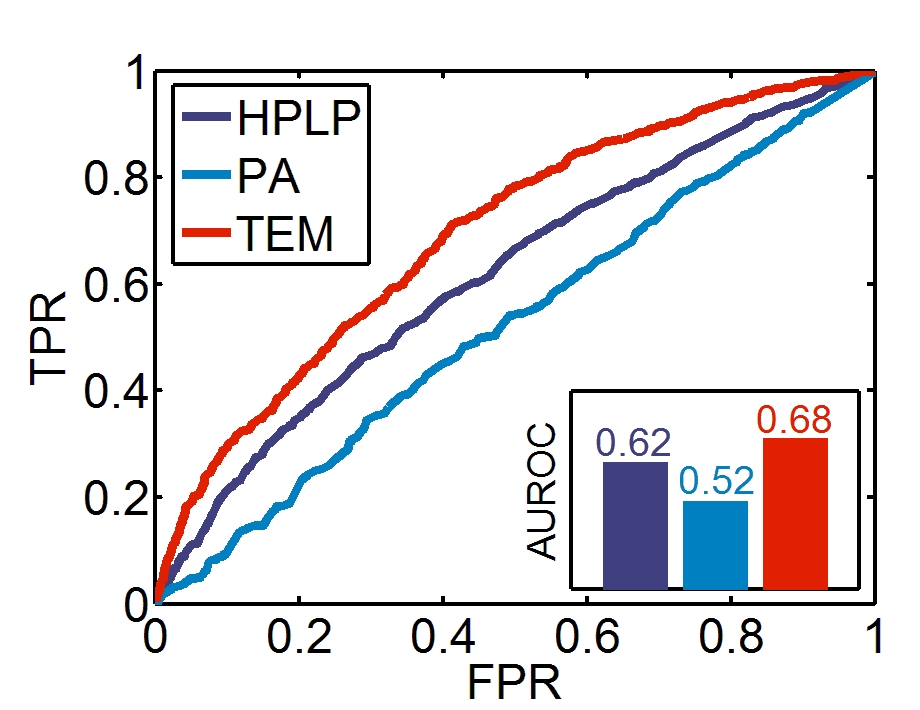}}\quad
	\subfloat[Enron-ROC]{\includegraphics[height=1.15in]{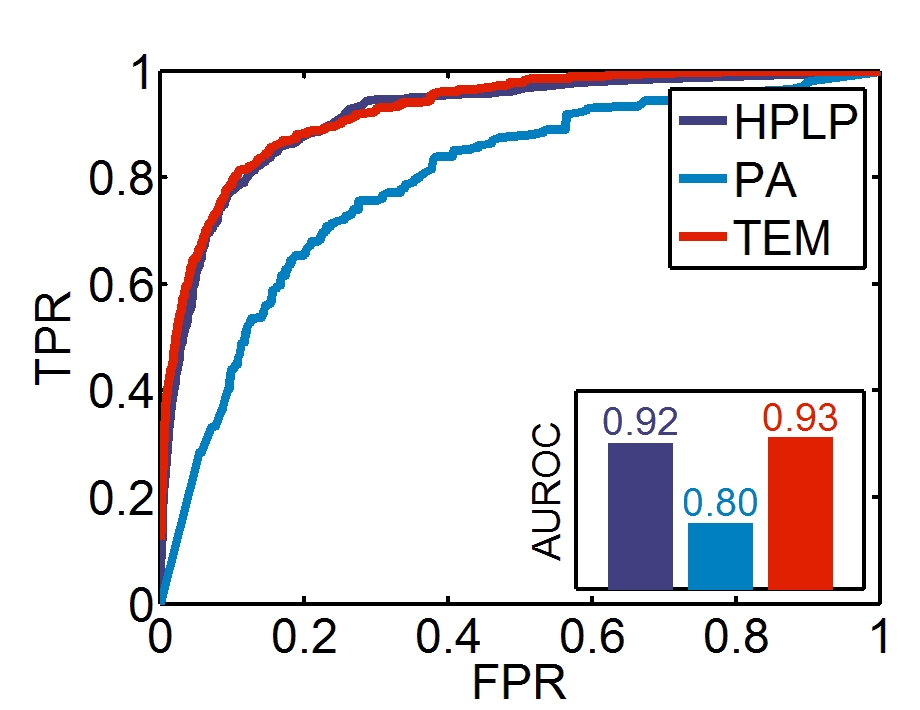}}\quad
	\subfloat[Facebook-ROC]{\includegraphics[height=1.15in]{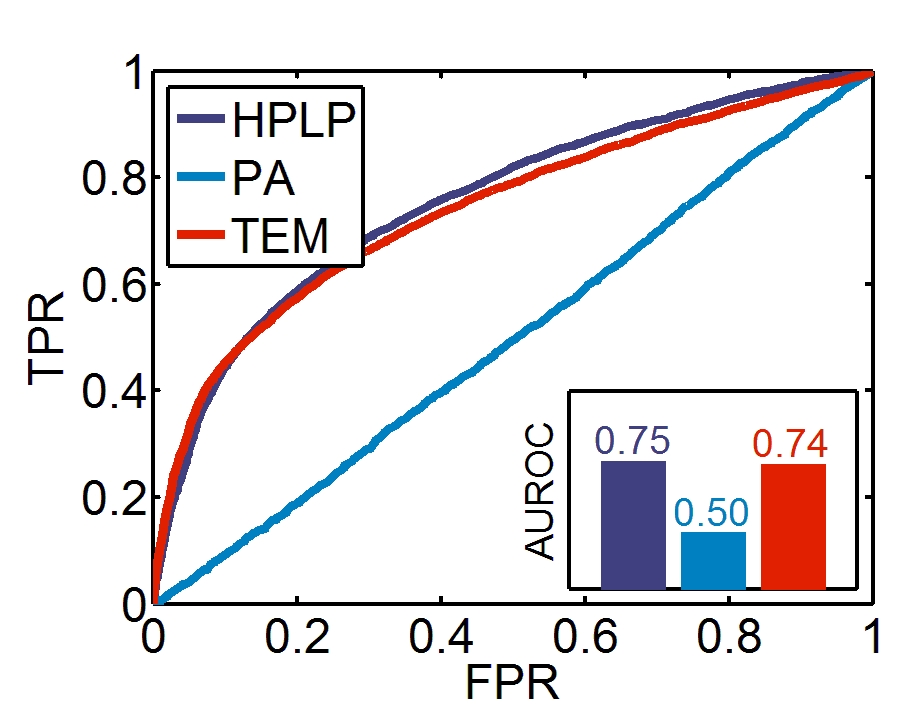}}\\
	\subfloat[Condmat-PR]{\includegraphics[height=1.15in]{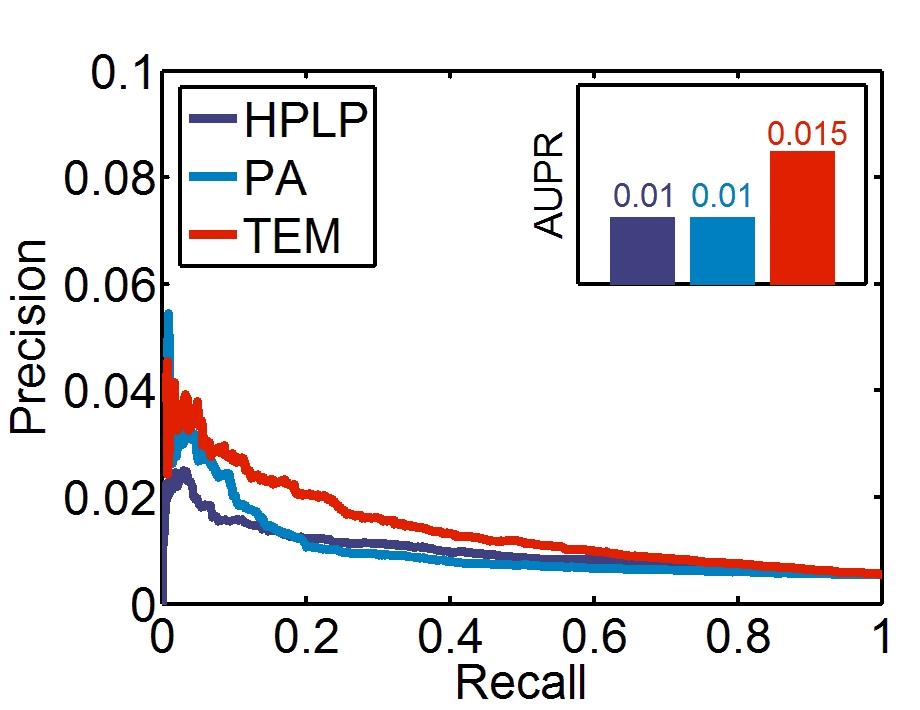}}\quad
	\subfloat[DBLP-PR]{\includegraphics[height=1.15in]{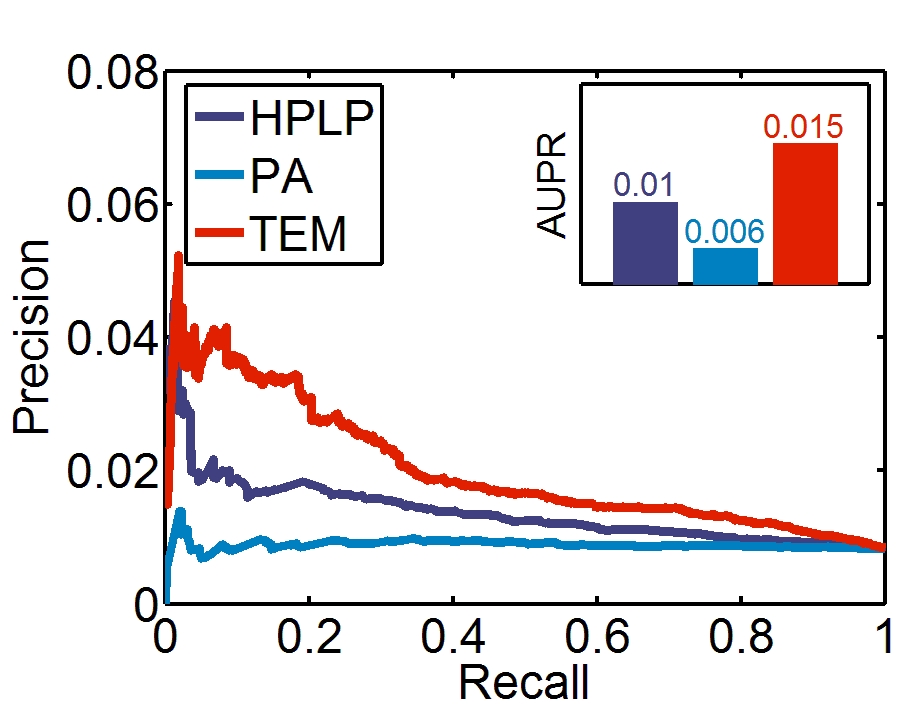}}\quad
	\subfloat[Enron-PR]{\includegraphics[height=1.15in]{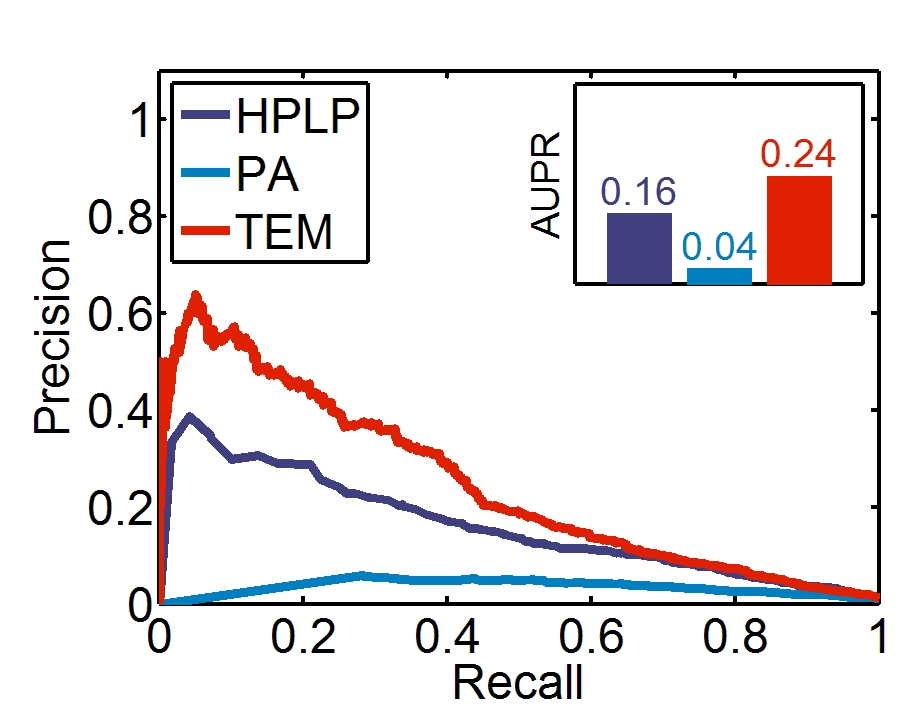}}\quad
	\subfloat[Facebook-PR]{\includegraphics[height=1.15in]{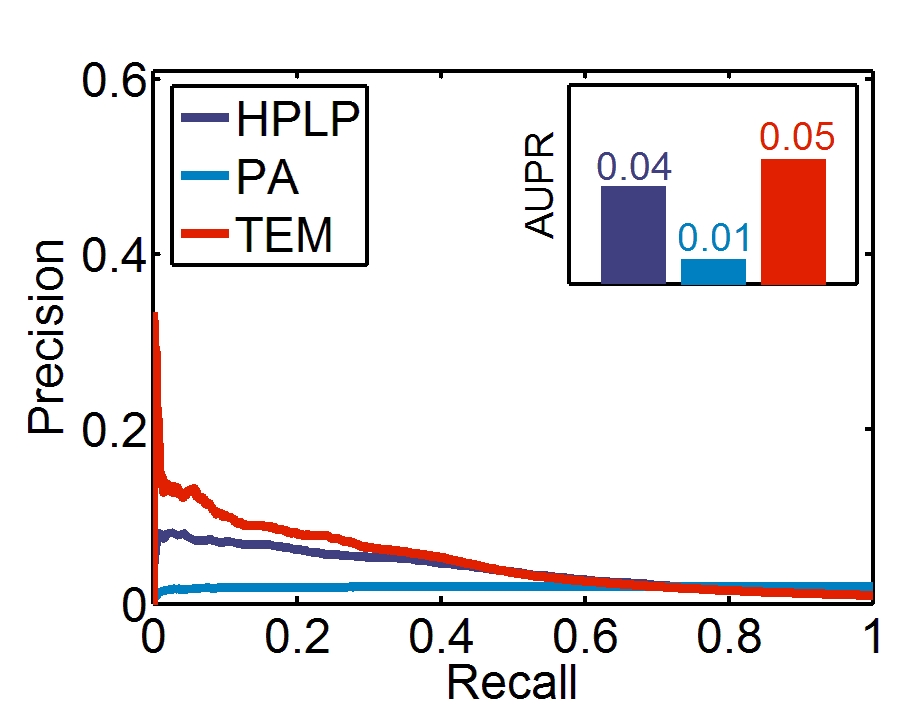}}\\
	\caption{Link Prediction Performance.}
\label{fig_lp_performance}
\end{figure*}

\begin{figure}[t]
\centering
\includegraphics[width=2.3in]{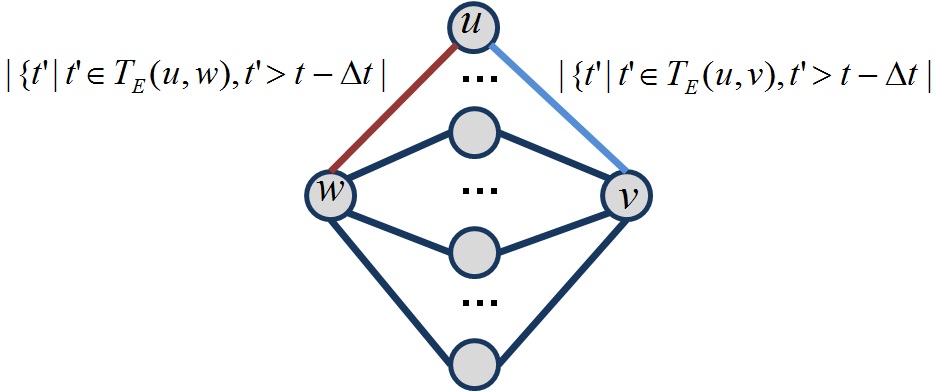}
\caption{Link Influence Probability Method}
\label{fig_triads_lp}
\end{figure}

\subsection{Inferring New Links}
{\bf Experimental Setup}
We set the default values for classifiers used in this paper, 10 bags of 10 random forest trees for HPLP (the same setting in the work of ~\cite{acm:linkprediction2}), 10 bags of 10 {\it logistic regression} for TEM-, TEM and TEM+. The features lists of all models are presented in Table~\ref{table_feature_list1}. {TEM} method combines $\text{TCE}_{s,t}$ and $\text{TEM}_{prob}(s,t)$ into the features vector, which includes the nodes pair collocation information and triad evolution information learned from historical data. In {TEM+} we include one more feature ({\it Link Influence Probability}) than the {TEM}, this feature is introduced in equation~\ref{eqn_lp}. While {TEM-} only includes the nodes pair collocation information described by $\text{TCE}_{s,t}$, where only static topological information are included. In this way we can investigate the generality of $\text{TEM}_{prob}(s,t)$ and {\it Link Influence Probability}. The performance of TEM- can be found in Table~\ref{table_general_across_lp}.

The reason we select {HPLP} for comparison is, HPLP includes almost all centrality measures frequently used in link analysis and it is also the best framework of feature-based link prediction till date. Another reason is, the HPLP method can be considered as a naive combination of preferential attachment and triadic closure, where the values of {preferential attachment} and {common neighbors} are simply combined into one feature vector. We undersample training set to 30\% positive class prevalence in training. We do not change the size or distribution of the testing data. In this paper, we restricted the prediction task within the set of two hops node pairs \cite{acm:linkprediction2}.

In Table~\ref{tab_lp_results} we present the performance comparisons of our methods with HPLP \cite{acm:linkprediction2} and state-of-art methods listed in \cite{acm:linkprediction1} ({\it Adamic/Adar}, {\it Common Neighbors} and {\it Preferential Attachment}). As suggested in the work \cite{acm:fairandeffective}, given the high class imbalance, area under the precision-recall curve (AUPR) should be used as the primary evaluation measure. We can see that the {TEM+} and {TEM} method significantly outperforms the {HPLP} method by at most 121\% in terms of AUPR. We observe almost the same pattern in Section~\ref{sec_infer_prominence}. Thus, the position profile methodology is consistently effective for both the prominence prediction and link prediction. Additionally we find that our framework (TEM+ and TEM) are better than PA, CN and AA methods. This implies that the combination of preferential attachment and triadic closure is better than each of them alone.

To be rigorous, in Figure~\ref{fig_lp_performance} we also provide ROC curves and PR curves for three methods, TEM, HPLP and PA. HPLP includes almost all classical predictors frequently used in link analysis, which can be considered as a naive combination of {\it preferential attachment} (i.e., PA) and {\it triadic closure} (i.e., common neighbors and Adamic/Adar). We observe that: 1) TEM method outperforms PA method significantly in terms of ROC curve and PR curve, which means {\it preferential attachment} is not the single origin underlying the process of link prediction; 2) HPLP method outperforms PA method in all cases, which indicates that both preferential attachment and triadic closure are not negligible in network evolution; 3) TEM method is better than HPLP method, which means TEM successfully optimizes certain trade-offs between {\it preferential attachment} and {\it triadic closure}. All of these results demonstrate the strength of our framework in predicting new links.
\section{Generalization across Datasets: A case for transfer learning}
\label{sec_generalization}
In the above sections we have demonstrated that the {\it triad position profile} has a stronger generalization capacity than nodal attributes based methods in predicting future prominent nodes and predicting new links. To be rigorous, we now ask: are these features powerful enough to transfer learning from one social network to another? If our framework are able to generalize across datasets, then it will further demonstrate that our framework captures the essential principles of network evolution.
\begin{table*}[t]
\caption{\bf{Generalization Measured in AUPR (Prominence Prediction)}}
\centering
\resizebox{6.9in}{!}{
\begin{tabular}{|c|c|c|c|c|c|c|c|c|c|c|c|c|c|c|} \hline
{\bf TPP} & Condmat & DBLP & Enron & Facebook & {\bf TPP+} & Condmat & DBLP & Enron & Facebook & {\bf Baseline} & Condmat & DBLP & Enron & Facebook \\ \hline \hline
Condmat & 0.716 & 0.171 & 0.450 & 0.191 & Condmat & 0.717 & 0.171 & 0.449 & 0.191 & Condmat& 0.712 & 0.153 & 0.327 & 0.151\\ \hline
DBLP & 0.632 & 0.221 & 0.406 & 0.341 & DBLP & 0.626 & 0.220 & 0.413 & 0.317 & DBLP & 0.525 & 0.189 & 0.270 & 0.248  \\ \hline
Enron & 0.681 & 0.269 & 0.516 & 0.415 & Enron & 0.671 & 0.293 & 0.519 & 0.421 & Enron & 0.650 & 0.270 & 0.506 & 0.429\\ \hline
Facebook & 0.617 & 0.310 & 0.480 & 0.426 & Facebook & 0.615 & 0.341 & 0.483 & 0.430 & Facebook & 0.359 & 0.320 & 0.344 & 0.417 \\ \hline 
\end{tabular}
}
\label{table_general_across}
\end{table*}
\begin{table*}[t]
\caption{\bf{Generalization Measured in AUPR (Link Prediction)}}
\centering
\resizebox{6.9in}{!}{
\begin{tabular}{|c|c|c|c|c|c|c|c|c|c|c|c|c|c|c|} \hline
{\bf TEM-} & Condmat & DBLP & Enron & Facebook & {\bf TEM} & Condmat & DBLP & Enron & Facebook & {\bf HPLP} & Condmat & DBLP & Enron & Facebook \\ \hline \hline
Condmat & 0.014 & 0.017 & 0.247 & 0.044 & Condmat & 0.015 & 0.021 & 0.226 & 0.049 & Condmat& 0.014 & 0.015 & 0.133 & 0.041\\ \hline
DBLP & 0.013 & 0.020 & 0.231 & 0.041 & DBLP & 0.013 & 0.020 & 0.240 & 0.042 & DBLP & 0.012 & 0.019 & 0.119 & 0.035  \\ \hline
Enron & 0.014 & 0.016 & 0.221 & 0.041 & Enron & 0.012 & 0.016 & 0.243 & 0.030 & Enron & 0.013 & 0.016 & 0.195 & 0.038\\ \hline
Facebook & 0.011 & 0.022 & 0.197 & 0.050 & Facebook & 0.012 & 0.022 & 0.165 & 0.050 & Facebook & 0.008 & 0.025 & 0.083 & 0.056 \\ \hline 
\end{tabular}
}
\label{table_general_across_lp}
\end{table*}
\subsection{Generalization-the Prominence Prediction}
We first consider the prominence prediction problem. In Table~\ref{table_general_across}, we provide the transferred learning results for {\it baseline} model and {\it TPP} model. Each pair of generalization is trained on the row dataset and evaluated on the column dataset by  Bagging with {\it logistic regression}, as before. The diagonal entries represent the performance of models which are trained and tested on the same dataset, which makes it convenient for comparisons. 

There are several observations. First, we find that few generalization entries have higher performance than their corresponding non-generalization entries (diagonal entries), for example two increased entries of {\it TPP} belong to the generalization from Enron and Facebook to DBLP. Second, we observe that the {\it TPP} model's performance degrades remarkably less than the {\it baseline} model in most cases. This indicates that the position profile of node captures principles that are more generic than the centrality based model, and this still holds even if the generalization is across different domains of networks. Third, the generalization of the position profile methodology is not significantly impacted by the fact that the difficulty of prediction is different across the network domains and the fact that the imbalance ratio between training set size and testing set size: for example, prominence prediction is more difficult on DBLP dataset, however the models learned on DBLP still have performance that are comparable to the corresponding diagonal entries; additionally even if DBLP training set size is only around 100, it can still work well on the testing sets of Condmat, Enron and Facebook, which are up to thousands. Fourth, the difficulty of prediction in each dataset is not affected by the generalization: all performances in the same column are in the same order of magnitude. Additionally, by comparing the performance of {\it TPP} and {\it TPP}+ we can see the {\it TPP} method is more stable in the generalization across datasets; the reason is trivial, the {\it TPP}+ contains two feature ({\it prominence\_prob} and {\it prominence\_index}) which are not that generic across different domains of datasets. This further confirms that the position profile is a general cross-domain property for the influence evolution analysis.

In conclusion, the position profile based model is notably more generic across different domains of networks, and the centrality based model is more particular to a specific dataset.

\subsection{Generalization-the Link Prediction}
As discussed before, the link formation and influence evolution are always accompanied with each other. We also conducted an empirical generalization of the link prediction task across different datasets in Table~\ref{table_general_across_lp}. To note that, in order to have fair comparison, we use 10 bags of {logistic regression} for all methods. Most of our observations made in the generalization of prominence prediction still hold for the link prediction problem. First, this validates the intimate interactions between the influence evolution and the link formation mechanisms; we see the same overall pattern. For example, both  {\it baseline} and {\it HPLP} have large drops in performance when generalizing from Facebook to Condmat and generalizing from Facebook to Enron. Second, this suggests that the properties captured by the position profile (in both problems) are indeed general across datasets.

In conclusion based on the generalization of the prominence prediction problem and the link prediction problem across datasets, we postulate that the positions where nodes are located are more significant in determining their evolution orbits than the nodal attributes possessed by them. {{Our methodology of position profile has a greater degree of precision than has heretofore been possible in depicting the network evolution.}} This is due to the optimized trade-offs between {\it triadic closure} and {\it preferential attachment} in our triad position profile methodology.
\section{Conclusions}
In this paper we analyzed several principles/mechanisms underlying the network evolution, mainly focused on two essential elements of the network evolution: the individual influence evolution and the link formation/dissolution mechanism. We demonstrated that position of a node in a local structure is strongly indicative of the influence progression or future prominence of the node in the social network. Building on this observation, we developed a prominence prediction method as well as a method for link prediction. We showed that the node prominence and the process of link formation are closely intertwined and impact the evolution of a network. We empirically demonstrated the improvement in performance over the baseline methods for both prominence prediction and link prediction across the four different datasets. We further established the generalization capacity of our methods under a transfer learning scenario --- we learned the classifier on one social network (using the proposed features) and tested on another social network. The performance trends clearly showed that our approach is able to capture essential properties or features underlying network evolution, which are general across different domains of social networks. 

These findings are important for several reasons. First, it provides microscopic evidence that the {\it triadic closure} is a fundamental principle underlying the social network evolution. Second, our methodology (triad position profile) is validated to optimize trade-offs between essential dimensions of network evolution ({\it preferential attachment} and {\it triadic closure}), then it is not surprising that, as a consequence, our approach yields accurate and generic performance in both microscopic problems. In summary, we have developed a new perspective for network evolution and developed a general purpose feature vector that can be used by different machine learning algorithms across different social networks.

\section*{Acknowledgment}
Research was sponsored by the Army Research Laboratory under Cooperative Agreement Number W911NF-09-2-0053, and by the grant FA9550-12-1-0405 from the U.S. Air
Force Office of Scientific Research (AFOSR) and the Defense
Advanced Research Projects Agency (DARPA).


%

\end{document}